\documentclass[aps,prb,twocolumn,floatfix,showpacs,superscriptaddress]{revtex4}
\usepackage[dvips]{color}
\usepackage{amsmath,amssymb,bm}
\usepackage{graphicx,dcolumn}
\usepackage[hypertex,breaklinks=true,colorlinks=false]{hyperref}

\newcommand{\chlab}{\nu}
\newcommand{\Xh}{\hat{X}}
\newcommand{\Ph}{\hat{P}}
\newcommand{\rvec}{\vec{r}}
\newcommand{\svec}{\vec{s}}
\newcommand{\tvec}{\vec{t}}
\newcommand{\uvec}{\vec{u}}
\newcommand{\vvec}{\vec{v}}
\newcommand{\uvech}{\hat{\uvec}}
\newcommand{\vvech}{\hat{\vvec}}

\newcommand{\avec}{\vec{a}}
\newcommand{\bvec}{\vec{b}}
\newcommand{\Phiv}{\vec{\Phi}}
\newcommand{\Thetav}{\vec{\Theta}}

\newcommand{\bra}[1]{\langle #1 |}
\newcommand{\ket}[1]{| #1 \rangle}
\newcommand{\bracket}[2]{\langle #1 | #2 \rangle}
\newcommand{\biggbra}[1]{\bigg\langle #1 \bigg|}
\newcommand{\biggket}[1]{\bigg| #1 \bigg\rangle}
\newcommand{\Dcal}{{\mathcal D}}
\newcommand{\Ecal}{{\cal E}}

\newcommand{\Ncal}{{\mathcal N}}

\newcommand{\Rcal}{{\mathcal R}}
\newcommand{\gammat}{\tilde{\gamma}}

\newcommand{\varphit}{\tilde{\varphi}}

\newcommand{\Ht}{\tilde{H}}
\newcommand{\nt}{\tilde{n}}
\newcommand{\Xit}{\tilde{\Xi}}
\newcommand{\qt}{\tilde{q}}
\newcommand{\rt}{\tilde{r}}
\newcommand{\Tr}{\mathrm{Tr}}
\newcommand{\PRL}[3]{Phys. Rev. Lett. {\bf #1},
\href{http://link.aps.org/abstract/PRL/v#1/e#2}{#2} (#3)}
\newcommand{\PRLp}[3]{Phys. Rev. Lett. {\bf #1},
\href{http://link.aps.org/abstract/PRL/v#1/p#2}{#2} (#3)}
\newcommand{\PRA}[3]{Phys. Rev. A {\bf #1},
\href{http://link.aps.org/abstract/PRA/v#1/e#2}{#2} (#3)}

\newcommand{\PRB}[3]{Phys. Rev. B {\bf #1},
\href{http://link.aps.org/abstract/PRB/v#1/e#2}{#2} (#3)}
\newcommand{\PRBp}[3]{Phys. Rev. B {\bf #1},
\href{http://link.aps.org/abstract/PRB/v#1/p#2}{#2} (#3)}
\newcommand{\PRBR}[3]{Phys. Rev. B {\bf #1},
\href{http://link.aps.org/abstract/PRB/v#1/e#2}{#2} (R) (#3)}

\newcommand{\RMP}[3]{Rev. Mod. Phys. {\bf #1}, 
\href{http://link.aps.org/doi/10.1103/RevModPhys.#1.#2}{#2} (#3)}
\newcommand{\arXiv}[1]{arXiv:\href{http://arxiv.org/abs/#1}{#1}}

\newcommand{\doilink}[2]{\href{http://dx.doi.org/#1}{#2}}

\begin{document}

%%%%%%%%%%%%%%%%%%%%%%%%%%%%%%%%%%%%%%%%%%%%%%%%%
% Paper Information
%%%%%%%%%%%%%%%%%%%%%%%%%%%%%%%%%%%%%%%%%%%%%%%%%

\title{Entanglement entropy between two coupled Tomonaga-Luttinger liquids}

\author{Shunsuke Furukawa}
\affiliation{Department of Physics, University of Toronto, Toronto, Ontario M5S 1A7, Canada}

\author{Yong Baek Kim}
\affiliation{Department of Physics, University of Toronto, Toronto, Ontario M5S 1A7, Canada}
\affiliation{School of Physics, Korea Institute for Advanced Study, Seoul 130-722, Korea}

%\email[]{furukawa@physics.utoronto.ca}
%\email[]{ybkim@physics.utoronto.ca}

\date{\today}

% 03.65.Ud  Entanglement and quantum nonlocality
% 03.67.Mn  Entanglement measures, witnesses, and other characterizations
% 05.50.+q  Lattice theory and statistics (Ising, Potts, etc.)
% 05.70.Jk  Critical point phenomena
% 11.25.Hf Conformal field theory 
% 64.60.-i  General studies of phase transitions
% 71.10.Pm  Fermions in reduced dimensions (anyons, composite fermions, Luttinger liquid, etc.) 
% 71.10.Hf 	Non-Fermi-liquid ground states, electron phase diagrams and phase transitions in model systems 
% 71.27.+a 	Strongly correlated electron systems; heavy fermions
% 75.10.Jm  Quantized spin models
% 75.10.Pq  Spin chain models
% 75.40.Mg  Numerical simulation studies

%\keywords{}

%%%%%%%%%%%%%%%%%%%%%%%%%%%%%%%%%%%%%%%%%%%%%%%%%
% Abstract
%%%%%%%%%%%%%%%%%%%%%%%%%%%%%%%%%%%%%%%%%%%%%%%%%
\begin{abstract}
We consider a system of two coupled Tomonaga-Luttinger liquids (TLL) on parallel chains 
and study the R\'enyi entanglement entropy $S_n$ {\it between} the two chains. Here the entanglement
{\it cut} is introduced between the chains, not along the perpendicular direction as used in previous
studies of one-dimensional systems.
The limit $n\to1$ corresponds to the von Neumann entanglement entropy.
The system is effectively described by two-component bosonic field theory 
with different TLL parameters in the symmetric/antisymmetric channels as far as
the coupled system remains in a gapless phase. 
We argue that in this system, $S_n$ is a linear function of the length of the chains (boundary law)
followed by a universal subleading constant $\gamma_n$ determined by the ratio of the two TLL parameters. 
The formulae of $\gamma_n$ for integer $n\ge 2$ 
are derived using (a) ground-state wave functionals of TLLs 
and (b) boundary conformal field theory, which lead to the same result. 
These predictions are checked in a numerical diagonalization analysis of a hard-core bosonic model 
on a ladder. 
Although the analytic continuation of $\gamma_n$ to $n\to 1$ turns out to be a difficult problem,
our numerical result suggests that the subleading constant in the von Neumann entropy is also universal. 
Our results may provide useful characterization of inherently anisotropic quantum phases such as the sliding Luttinger liquid
phase via qualitatively different behaviors of the entanglement entropy with the entanglement partitions along
different directions.
\end{abstract}
\maketitle

%%%%%%%%%%%%%%%%%%%%%%%%%%%%%%%%%%%%%%%%%%%%%%%%%
% Main text
%%%%%%%%%%%%%%%%%%%%%%%%%%%%%%%%%%%%%%%%%%%%%%%%%

%************************************************
\section{Introduction}
%************************************************

%-----------------------------
%- Introduction - TLL 
The concept of Tomonaga-Luttinger liquid (TLL)
provides a universal framework for studying various one-dimensional (1D) interacting systems.\cite{Giamarchi04} 
The low-lying excitations of such a system, either fermionic\cite{Haldane81_JPhysC} or bosonic,\cite{Haldane81_PRL} are essentially collective, 
and can be recast into a bosonic field theory describing the density and phase fluctuations. 
A spinless TLL is characterized by a continuously varying parameter $K$ (so-called TLL parameter), 
which appears in the exponents of correlation functions in the ground state 
and experimentally in the power-law temperature dependence of response functions. 
When two spinless TLLs are coupled (or when an interaction is introduced in a 1D gas of spin-$\frac12$ particles), 
the bosonic fields are reorganized into symmetric and antisymmetric channels, 
which can independently form TLLs. 
This is a fundamental mechanism which also underlies the spin-charge separation in a 1D electron gas. 
Interestingly, this idea has been generalized to a two-dimensional (2D) array of coupled TLLs,  
predicting a novel non-Fermi liquid phase, called sliding Luttinger liquid, 
which shows highly anisotropic correlations.\cite{Schulz83,Emery00,Vishwanath01}
A fundamental question related to these studies is 
in what way the system of coupled TLLs are distinguished from more conventional phases 
such as Fermi liquids or from the decoupled TLLs. 
Stimulated by the recent advances in applying quantum information tools to many-body systems, 
we here address this question using one of such tools ---
the entanglement entropy in the ground-state wave function. 

%----------------------------
%- Introduction - entanglement entropy

By partitioning the system into a subregion $A$ and its complement $\bar{A}$, 
the entanglement entropy is defined as the von Neumann entropy $S_A= - \Tr \rho_A \ln \rho_A$ 
of the reduced density matrix $\rho_A = \Tr_{\bar A} \ket{\Psi}\bra{\Psi}$, 
where $\ket{\Psi}$ is the ground state of the system. 
When the system contains only short-range correlations, 
$A$ and $\bar{A}$ correlate only in the vicinity of the boundary, 
and the entanglement entropy scales with the size of the boundary (boundary law).\cite{Srednicki93,Eisert10}  
Deviation from the boundary law signals the presence of certain non-trivial correlations, 
and furthermore can contain universal numbers characterizing the system. 
In one-dimensional critical systems, for example, 
the entanglement entropy $S_A$ for an interval embedded in the system 
shows a logarithmic scaling, 
whose coefficient reveals the central charge $c$ of 
the underlying conformal field theory (CFT).
\cite{Holzhey94,Vidal03,Calabrese04,Ryu06}
Possible further information of CFT such as the TLL parameter $K$ 
can be encoded in a multi-interval entanglement entropy\cite{Casini04,FPS09,Calabrese09} 
and in corrections to the universal scalings.\cite{Laflorencie06,Calabrese10}. 
In topologically ordered systems\cite{Kitaev06,Levin06,Hamma05}  
and in some 2D critical systems,\cite{Metlitski09,Hsu09,Stephan09,Hsu10,Oshikawa10,Stephan10} 
the entanglement entropy obeys a boundary law, 
but there appears a subleading universal constant which is determined from the basic properties of the ground state. 

%----------------------------
%- This paper
In this paper, 
we aim to characterize the quantum entanglement arising from the coupling of TLLs. 
We consider, as the simplest situation, a system of two coupled spinless TLLs defined on parallel periodic chains (rings),
and study the entanglement entropy between the two chains. 
The system is effectively described by a two-component bosonic field theory 
with different TLL parameters $K_\pm$ in the symmetric and antisymmetric channels. 
If we identify the two chains with the spin-$\frac12$ degrees of freedom, 
these channels correspond to the charge and spin modes, respectively. 
For 1D systems, the entanglement entropy has so far been studied mostly 
for an interval embedded in the chain, which can count the central charge in critical systems. 
We here instead partition the system into two rings. 
We expect that this partitioning is more useful in observing the effects of the coupling of the two TLLs. 
Furthermore, we expect that the present setting provides a good starting point 
for understanding possibly highly anisotropic characters of entanglement in a 2D sliding Luttinger liquid. 

%----------------------------
%- Renyi entropy
Specifically, we construct the reduced density matrix $\rho_A$ for one of the chains by tracing out the other, 
and compute the R\'enyi entanglement entropy: 
\begin{equation}\label{eq:Renyi}
 S_n = \frac{-1}{n-1} \log (\Tr ~\rho_A^n). 
\end{equation}
The limit $n\to 1$ corresponds to the von Neumann entanglement entropy: 
\begin{equation}
 S_1 \equiv \lim_{n\to 1} S_n = -\Tr ~\rho_A \log \rho_A. 
\end{equation}
The limit $n\to \infty$ corresponds to the so-called single-copy entanglement\cite{Eisert05}: 
\begin{equation}
 S_\infty = -\log \lambda_{\rm max}, 
\end{equation}
where $\lambda_{\rm max}$ is the largest eigenvalue of $\rho_A$. 
It is known that the values of $S_n$ with integer $n\ge 2$ 
determine the full eigenvalue distribution of $\rho_A$ 
(so-called entanglement spectrum).\cite{Calabrese08} 
When there is no coupling between the chains, $S_n$ is simply equal to zero. 
The entropy $S_n$ increases as the coupling increases.

%----------------------------
%- Main result
We will see that $S_n$ (with $n=1,2,\dots,\infty$) obeys a linear function of the chain length $L$:
\begin{equation}\label{eq:Sn_L}
 S_n = \alpha_n L + \gamma_n + \dots, 
\end{equation}
where the ellipsis represents terms which are negligible in the limit $L\to \infty$. 
The first term $\alpha_n L$ can be simply viewed as a boundary law contribution, 
and the coefficient $\alpha_n$ depends on microscopic details. 
Our main interest lies in the subleading constant $\gamma_n$. 
We argue that this constant is universal 
and is determined by the ratio of two TLL parameters, $K_+/K_-$. 

Recently, Poilblanc\cite{Poilblanc10} studied the entanglement entropy for a similar partitioning 
in {\it gapped} phases of a spin ladder model. 
In his results, the entanglement entropy shows a similar linear scaling, 
but a subleading constant was not identified. 
We expect that the linear scaling is a generic feature of this type of partitioning, 
and that the appearance of the subleading constant is characteristic of critical systems. 

%----------------------------
%- Organization of the paper

The paper is organized as follows. 
In Sec.~\ref{sec:setting}, we set up the problem which we consider in this paper, 
and present path integral representations of the reduced density matrix moments $\Tr~\rho_A^n$ (with integer $n\ge 2$). 
Based on these representations, in Secs.~\ref{sec:wavefn} and \ref{sec:BCFT}, 
we calculate the moments using two different approaches. 
In Sec.~\ref{sec:wavefn}, we use the field theoretical representations of the TLL ground-state wave functions. 
In Sec.~\ref{sec:BCFT}, we use a modern technique in boundary CFT based on boundary states and compactification lattices. 
The two approaches are complementary: 
while the derivation is simpler in the former, 
the latter provides a more systematic treatment which does not require any regularization procedure. 
Both the approaches lead to the linear scaling of $S_n$  
and the same formulae for the subleading constant $\gamma_n$. 
The expressions of $\gamma_n$ (as a function of $K_+/K_-$) 
are summarized in Sec.~\ref{sec:Renyi_ent_expressions}, 
together with a discussion on their analytic properties. 
In particular, we discuss a difficulty in analytically continuing the formulae of $\gamma_n$ (obtained for integer $n\ge 2$) 
to the von Neumann limit $n\to 1^+$. 
%In particular, it is found that the formulae of $\gamma_n$ (obtained for integer $n\ge 2$) 
%are not analytic in the limit $n\to 1^+$. 
In Sec.~\ref{sec:numerics}, we check our predictions on $S_n~(n\ge 2)$ in a numerical diagonalization analysis of a hard-core bosonic model on a ladder. 
While we do not have any analytic prediction on the von Neumann entropy $S_1$, 
the numerical result suggests that $S_1$ obeys a linear scaling as the R\'enyi entropies does 
and that the subleading constant $\gamma_1$ is universal. 
We conclude with a summary in Sec.~\ref{sec:conclusions}. 
Implications of our results on a 2D sliding Luttinger liquid 
are also presented. 

%************************************************
\section{Setup of the problem}  \label{sec:setting}
%************************************************

In this section, we set up the system and the problem which we consider in this paper. 
In particular, we present the path integral representations of the reduced density matrix moments $\Tr~\rho_A^n$ (with integer $n\ge 2$), 
which will be used in the following sections. 

%************************************************
\subsection{Coupled Tomonaga-Luttinger liquids} \label{sec:CTLL}
%************************************************

%----------------------------
%- Two TLLs
We consider a system of two TLLs $H_\nu~(\nu=1,2)$ on parallel periodic chains of length $L$ 
coupled via interactions $H_{12}$. 
We assume that the two TLLs are equivalent 
and are described by the Gaussian Hamiltonian: 
\begin{equation}\label{eq:H_1_2}
 H_\chlab = \int_0^L dx ~\frac{v}2 \left[ 
 K\left(\frac{d\theta_\chlab}{dx}\right)^2 + \frac1K \left(\frac{d\phi_\chlab}{dx}\right)^2
 \right], \quad
 \chlab=1,2, 
\end{equation}
where $x$ is the coordinate along the chains, 
and $v$ and $K$ are the velocity and the TLL parameter, respectively, in each chain. 
In the case of fermions, $K<1$ ($K>1$) corresponds to a repulsive (attractive) intra-chain interaction. 
The dual pair of bosonic fields, $\phi_\chlab$ and $\theta_\chlab$, satisfy 
$[\phi_\chlab(x),\theta_{\chlab'}(x')] = (i/2) [1+\mathrm{sgn} (x-x')] \delta_{\chlab\chlab'}$. 
The field $\phi_\chlab (x)$ is related to the particle density $\rho_\chlab(x)$ 
via $\rho_\chlab(x)\approx \rho_0 - \frac{1}{\sqrt{\pi}}\frac{d\phi_\chlab(x)}{dx}$ 
with $\rho_0$ being the density in the ground state 
while the field $\theta_\chlab(x)$ represents the Josephson phase.  
We assume that there is no particle tunneling between the chains, 
and therefore the particle number is separately conserved in each chain ($U(1)\times U(1)$ symmetry). 

%----------------------------
%- Coupling between TLLs
Now let us consider, for instance, the interaction of the form 
\begin{equation}
 H_{12} = \int_0^L dx~ \frac{U}\pi \frac{d\phi_1}{dx} \frac{d\phi_2}{dx}, 
\end{equation}
which corresponds to the leading part in the density-density interaction. 
To treat this, 
we introduce the symmetric/antisymmetric combinations of the bosonic fields:
\begin{equation}\label{eq:boson+-}
 \phi_\pm = \frac1{\sqrt{2}} (\phi_1\pm\phi_2), \quad
 \theta_\pm = \frac1{\sqrt{2}} (\theta_1\pm\theta_2). 
\end{equation}
Then the total Hamiltonian $H=H_1+H_2+H_{12}$ can be formally decoupled into two free bosons 
defined for these symmetric/antisymmetric channels: 
\begin{equation}\label{eq:H+H-}
 H=H_+ + H_-, 
\end{equation}
with 
\begin{equation}\label{eq:H+-}
 H_\pm = \int_0^L dx ~\frac{v_\pm}2 \left[ 
 K_\pm \left(\frac{d\theta_\pm}{dx}\right)^2 + \frac1{K_\pm} \left(\frac{d\phi_\pm}{dx}\right)^2
 \right].  
\end{equation}
Here the renormalised velocities $v_\pm$ and TLL parameters $K_\pm$ are given by 
\begin{equation}\label{eq:vpm_Kpm}
 v_\pm = v \left( 1\pm \frac{KU}{\pi v} \right)^{\frac12}, 
 \quad 
 K_\pm = K \left( 1\pm \frac{KU}{\pi v} \right)^{-\frac12}. 
\end{equation}
Note that, although the two channels are formally decoupled in Eq.~\eqref{eq:H+H-}, 
zero modes of the two channels are intertwined, 
which will be seriously discussed in Sec.~\ref{sec:BCFT}. 
On the other hand, the oscillator modes of these channels are completely decoupled. 
In general, if $H_{12}$ consists only of forward scattering processes, 
the total Hamiltonian $H$ can be similarly recast into the form in Eqs.~\eqref{eq:H+H-} and \eqref{eq:H+-}. 
Even when $H_{12}$ contains other terms, this form is still applicable  
as long as those terms are irrelevant and diminish to zero in the renormalization group (RG) flow. 
In this case, $K_\pm$ can change slightly from the perturbative result [like Eq.~\eqref{eq:vpm_Kpm}] along the RG flow, 
and their precise values in the infra-red limit can be determined by examining correlation functions numerically, for example. 
In the following, we consider the situation where the Hamiltonian in Eqs.~\eqref{eq:H+H-} and \eqref{eq:H+-} 
presents the exact long-distance physics,  
and treat $K_\pm$ as free parameters.  
Since we are interested in the entanglement between the two chains, 
we keep in mind that the bosonic fields $\phi_\pm$ and $\theta_\pm$ diagonalizing $H$ 
are related to the original fields on the chains via Eq.~\eqref{eq:boson+-}. 
Note that Eq.~\eqref{eq:boson+-} is protected by the permutation symmetry of the two chains 
and is applicable beyond the perturbative regime.

%Note that in this case $K_\pm$ achieves further modification along the RG flow. 
%In the infrared limit, the Hamiltonian in Eqs.~\eqref{eq:H+H-} and \eqref{eq:H+-} then presents an exact description of the long-distance physics. 
%In the following, we use Eqs.~\eqref{eq:boson+-}, \eqref{eq:H+H-} and \eqref{eq:H+-} as the starting point of the discussions 

% Because the total Hamiltonian $H=H_1+H_2+H_{12}$ is symmetric with respect to the permutation of the two chains, 
% it is useful to introduce 
%Under certain condition 
%(more specifically, when the interaction $H_{12}$ consists of forward scattering and irrelevant perturbations), 

%************************************************
\subsection{Path integral representations of reduced density matrix moments} \label{sec:path_rdm}
%************************************************

%############################
\begin{figure}
\includegraphics[width=0.5\textwidth]{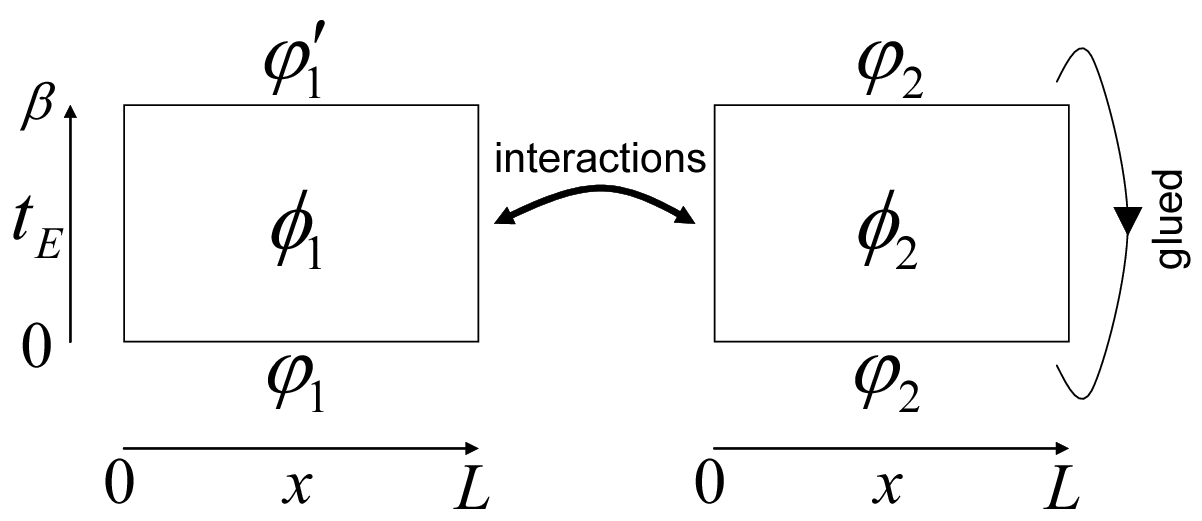}%{rdm.eps}
\caption{
The path integral representation for the reduced density matrix $\rho_A$ in Eq.~\eqref{eq:rdm_path}. 
In each of the two sheets, a periodic boundary condition is imposed in the $x$ direction. 
Therefore the left and right sheets form a cylinder and a torus, respectively. 
}
\label{fig:rdm}
\end{figure}
%############################

%----------------------------
%- Path integral
For the ground state $\ket{\Psi}$ of $H$, 
we consider the R\'enyi entanglement entropy $S_n$ [Eq.~\eqref{eq:Renyi}] with integer $n\ge 2$ between the two chains. 
Here we represent the moments of the reduced density matrix, $\Tr~ \rho_A^n$, in the language of the path integral. 
We start from the finite-temperature density matrix of the total system:
\begin{equation}
 \rho = \frac1Z  e^{-\beta H} 
 ~~\text{with}~~
 Z=\Tr~ e^{-\beta H}.
\end{equation}
The inverse temperature $\beta$ is eventually taken to infinity so that $\rho \to \ket{\Psi}\bra{\Psi}$. 
We move on to the path integral formalism in the Euclidean space time $(t_E,x)$.  
The Euclidean action is 
\begin{equation}
 S_E = \int_0^\beta dt_E \int_0^L dx ~ ( {\cal L}_{E+} + {\cal L}_{E-} )
\end{equation}
with 
\begin{equation}
 {\cal L}_{E\pm} = \frac{v_\pm}{K_\pm} \left[ (\partial_x \phi_\pm)^2 + v_\pm^{-2} (\partial_{t_E} \phi_\pm)^2 \right]. 
\end{equation}
Although $S_E$ is diagonalized in $\phi_\pm$ basis, in the following, 
we rather regard this as a functional of $\phi_{1,2}$ using the relation \eqref{eq:boson+-} 
since we are interested in the entanglement between the two chains. 
On this ground, the matrix element of the density matrix $\rho$ is expressed as
\begin{equation}\label{eq:rho_el}
\begin{split}
 &\bra{\varphi_1',\varphi_2'} \rho \ket{\varphi_1,\varphi_2} \\
 &=\frac1Z \int_
   {\footnotesize \begin{matrix}
   \phi_\chlab(0,x)=\varphi_\chlab(x)\\
   \phi_\chlab(\beta,x)=\varphi_\chlab'(x)
   \end{matrix}} 
  \Dcal\phi_1 \Dcal\phi_2 ~e^{-S_E[\phi_1,\phi_2]}, 
\end{split}
\end{equation}
where $\varphi_\chlab=\{\varphi_\chlab(x) \}_{0\le x <L}$ and those with a prime are field configurations defined along the chains 1 and 2 respectively. 
The path integral is done under the condition that $\phi_\chlab(t_E,x)~(\chlab=1,2)$ is equal to 
$\varphi_\chlab(x)$ and $\varphi_\chlab'(x)$ at the imaginary time $t_E=0$ and $\beta$, respectively.

%----------------------------
%- Reduced density matrix
The reduced density matrix $\rho_A$ for the chain $1$ is obtained 
by identifying $\varphi_2$ and $\varphi_2'$ in Eq.~\eqref{eq:rho_el} and integrating over $\varphi_2$:
\begin{equation}\label{eq:rdm_path}
\begin{split}
 &\bra{\varphi_1'} \rho_A \ket{\varphi_1} 
 = \int \Dcal \varphi_2 ~ \bra{\varphi_1', \varphi_2} \rho \ket{\varphi_1, \varphi_2}\\
 &=\frac1Z \int_
   {\footnotesize \begin{matrix}
   \phi_1(0,x)=\varphi_1(x)\\
   \phi_1(\beta,x)=\varphi_1'(x)\\
   \phi_2(0,x)=\phi_2(\beta,x)
   \end{matrix}} 
  \Dcal\phi_1 \Dcal\phi_2 ~e^{-S_E[\phi_1,\phi_2]}, 
\end{split}
\end{equation}
with $\Dcal\varphi_\chlab = \prod_x d \varphi_\chlab(x)$. 
We introduce a graphical representation in Fig.~\ref{fig:rdm}, 
where two sheets express the space-time on which the fields $\phi_{1,2}$ are defined. 
The partial trace in Eq.~\eqref{eq:rdm_path}
corresponds to gluing the two edges of the sheet for $\phi_2$.

%############################
\begin{figure}
\includegraphics[width=0.5\textwidth]{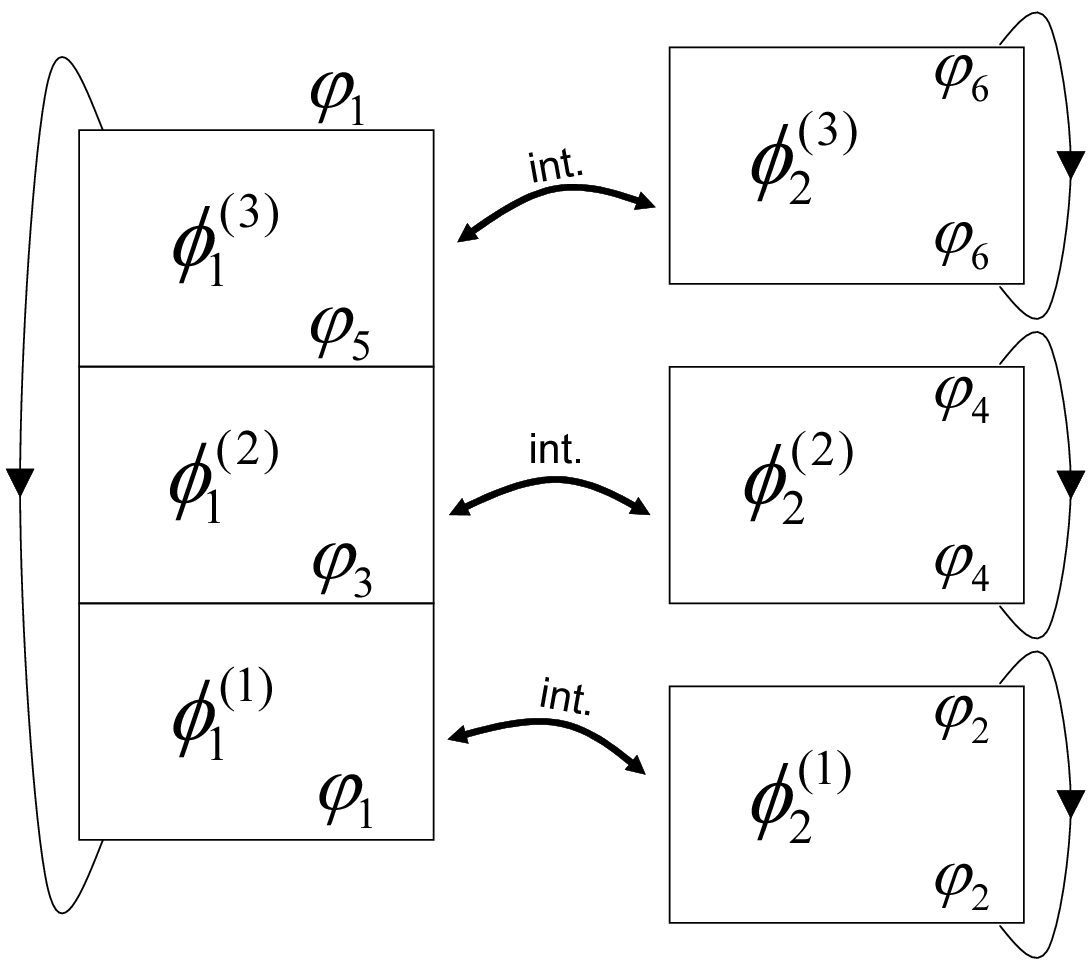}%{Z3.eps}
\caption{
The path integral representation of $\Tr \rho_A^n$ in Eqs.~\eqref{eq:rho_n} and \eqref{eq:rho_n_Zn}, 
for the case of $n=3$. 
}
\label{fig:Z3}
\end{figure}
%############################

%----------------------------
%- Moment of the reduced density matrix
Now we consider the $n$-th moment of the reduced density matrix, $\Tr \rho_A^n$, 
with integer $n\ge 2$. 
To construct this, we consider $n$ copies of the diagram in Fig.~\ref{fig:rdm} 
and glue them cyclically, 
as illustrated in Fig~\ref{fig:Z3} for the case of $n=3$. 
This leads to an expression 
\begin{equation}\label{eq:rho_n}
\begin{split}
 \Tr ~ \rho_A^n 
 &= \int \prod_{j=1}^n \Dcal\varphi_{2j-1} ~ \prod_{j=1}^n \bra{\varphi_{2j+1}} \rho_A \ket{\varphi_{2j-1}}\\
 &= \int \prod_{j=1}^{2n} \Dcal\varphi_j ~ \prod_{j=1}^n \bra{\varphi_{2j+1},\varphi_{2j}} \rho \ket{\varphi_{2j-1},\varphi_{2j}}, 
\end{split}
\end{equation}
where $\varphi_j$'s with odd (even) subscripts are defined for the chain 1 (2) 
and $\varphi_{2n+1}\equiv \varphi_1$. 
This can be expressed in a compact way:
\begin{equation}\label{eq:rho_n_Zn}
 \Tr \rho_A^n = \frac{Z_n}{Z^n}, 
\end{equation}
where $Z_n$ is the partition function defined for $2n$ sheets  
which are interconnected as shown in Fig.~\ref{fig:Z3}. 
The diagram consists of a large torus for $\phi_1$'s, 
and $n$ small tori for $\phi_2$'s. 
Because of the interactions between $\phi_1$ and $\phi_2$, 
the calculation of such a partition function is not trivial. 
In Secs.~\ref{sec:wavefn} and \ref{sec:BCFT}, 
we present different ways to compute Eq.~\eqref{eq:rho_n} or Eq.~\eqref{eq:rho_n_Zn}, 
which eventually lead to the same result. 
Here we mention the case of no inter-chain interaction $H_{12}=0$, 
where the tori in Fig.~\ref{fig:Z3} are decoupled. 
Using the ground-state energy $E_0$ of $H_1$ and $H_2$, we have 
$Z_n \approx e^{-n\beta E_0} \left( e^{-\beta E_0} \right)^n$ and $Z^n \approx \left(e^{-2\beta E_0}\right)^n$ 
in the limit $\beta\to \infty$, 
which lead to $S_n=0$.

%************************************************
\section{Wave functional approach} \label{sec:wavefn}
%************************************************

%----------------------------
%- Section introduction
In this section we compute Eq.~\eqref{eq:rho_n} 
using a field theoretical representation of the TLL wave function, 
and derive the expressions of the R\'enyi entropies $S_n$ for integer $n\ge 2$. 
Similar approaches were also used 
to calculate the entanglement entropy in 2D critical wave functions\cite{Stephan09}
and the ground-state fidelity in TLLs.\cite{Yang07,Fjaerestad08,CamposVenuti09} 
In this section, we do not include the zero modes of the bosonic fields 
and regard $H_\pm$ as completely independent. 
This is justified because we are interested in the entanglement properties of the ground state, 
where zero modes do not appear. 
% A more careful treatment of the zero modes is presented in Sec.~\ref{sec:BCFT}, 
% but a simple calculation in this section focusing on the oscillator modes 
% finally produces the same result. 

%************************************************
\subsection{Reduced density matrix moments and wave functionals}
%************************************************

%############################
\begin{figure}
\includegraphics[width=0.45\textwidth]{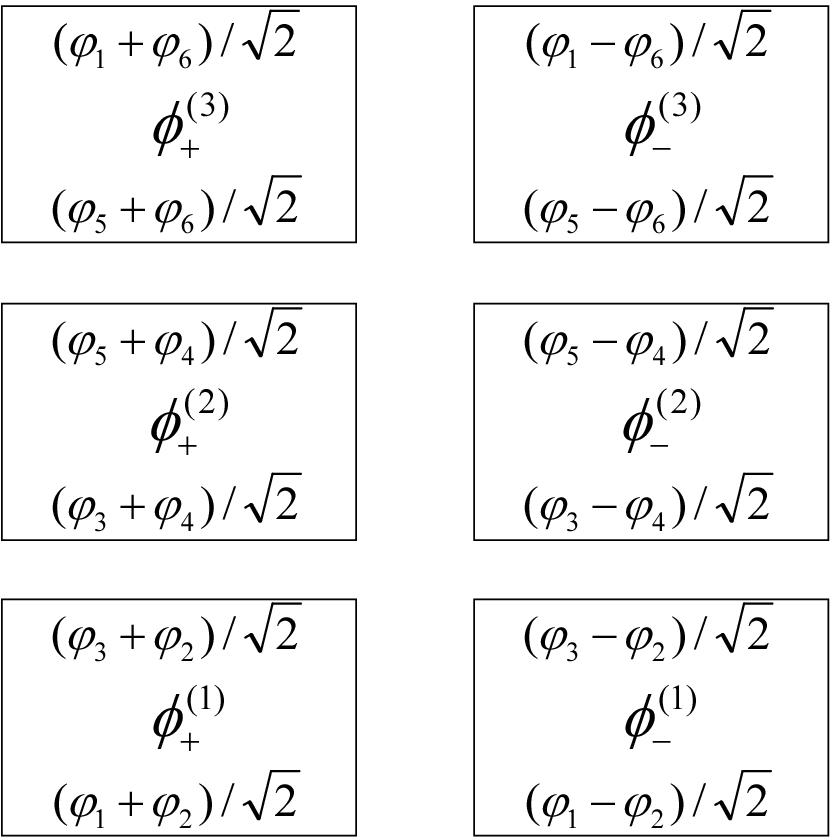}%{Z3_pm.eps}
\caption{
Rewriting of $Z_3$ in Fig.~\ref{fig:Z3} using $\phi_\pm$ basis. 
The partition function $Z_3$ is obtained after integrating over the field configurations $\varphi_1,\dots,\varphi_6$. 
}
\label{fig:Z3_pm}
\end{figure}
%############################

%----------------------------
%- Transformation to +- channels
The difficulty in computing $Z_n$ comes from the interactions between different sheets in Fig.~\ref{fig:Z3}. 
To treat these interactions,  
we work in the symmetric/antisymmetric basis of bosonic fields, 
in which the action is diagonal, leading to a diagram as in Fig.~\ref{fig:Z3_pm}. 
As a tradeoff, the boundaries of the sheets are now interconnected in a non-trivial way. 
The strategy of this section is to first treat each sheet of Fig.~\ref{fig:Z3_pm} separately 
by fixing the boundary field configurations, $\varphi_j$'s, 
and to then integrate over $\varphi_j$'s to calculate the partition function $Z_n$. 

% The strategy of this section for computing Eq.~\eqref{eq:rho_n} or Eq.~\eqref{eq:rho_n_Zn} is state as follows. 
% The difficulty in comes $Z_n$ comes from the interactions between different sheets in Fig.~\ref{fig:Z3}. 
% To treat these interactions,  
% we first fix the field configurations $\varphi_j$ ($j=1,\dots,2n$) along the boundaries.  
% We then move on to the symmetric/antisymmetric basis of bosonic fields, 
% in which the action is diagonal, leading to a diagram as in Fig.~\ref{fig:Z3_pm}. 
% We treat each sheet of Fig.~\ref{fig:Z3_pm} separately by fixing the boundary field configurations 
% and then integrate over $\varphi_j$'s to calculate the partition function $Z_n$. 

%----------------------------
%- Expression using string propagator
As mentioned in Sec.~\ref{sec:CTLL}, the winding numbers (zero modes) of the symmetric/antisymmetric channels are intertwined, 
and therefore the two channels are not completely decoupled. 
However, since we are interested in the entanglement properties of the ground state in the limit $\beta\to\infty$, 
we can work in the sector of the Hilbert space where the winding numbers are set to zero. 
Namely, we focus on the oscillator modes in the Hamiltonian. 
In this sector, $H_\pm$ commute with each other. 
Using $e^{-\beta H}=e^{-\beta H_+} e^{-\beta H_-}$, we rewrite the matrix element of $\rho$ appearing in Eq.~\eqref{eq:rho_n} as 
\begin{equation}\label{eq:rho_el_pm}
\begin{split}
 \bra{\varphi_{2j+1}&,\varphi_{2j}} \rho \ket{\varphi_{2j-1},\varphi_{2j}} \\
 =\frac1Z 
 &\biggbra{ \frac{\varphi_{2j+1}+\varphi_{2j}}{\sqrt{2}} } e^{-\beta H_+} \biggket{ \frac{\varphi_{2j-1}+\varphi_{2j}}{\sqrt{2}} }\\
 \times 
 &\biggbra{ \frac{\varphi_{2j+1}-\varphi_{2j}}{\sqrt{2}} } e^{-\beta H_-} \biggket{ \frac{\varphi_{2j-1}-\varphi_{2j}}{\sqrt{2}} }. 
\end{split}
\end{equation}
An expression of the form $\bra{\varphi'} e^{-\beta H_\pm} \ket{\varphi}$ in this equation 
corresponds to each sheet in Fig.~\ref{fig:Z3_pm}. 
Since $H_\pm$ are the Hamiltonians of massless free bosons, 
$\bra{\varphi'} e^{-\beta H_\pm} \ket{\varphi}$ can be viewed 
as the propagator of a closed bosonic string in the imaginary time. 
Such a ``closed string propagator'' has been computed in, e.g., Refs.~\onlinecite{Cohen86,Lorenzo86,Mezincescu89}
[in particular a compact expression is shown in Eq.~(24) of Ref.~\onlinecite{Mezincescu89}]. 
Rather than using the expression obtained in these works, 
we here take a simpler route as follows. 

%----------------------------
%- wave functional
We take the limit $\beta\to\infty$, 
and then only the ground states $\ket{\Psi_\pm}$ of $H_\pm$ (with eigenenergies $E_\pm$) 
contribute to the propagators and the partition function: 
\begin{align}
 &\bra{\varphi'} e^{-\beta H_\pm} \ket{\varphi} \approx \bracket{\varphi'}{\Psi_\pm} e^{-\beta E_\pm} \bracket{\Psi_\pm}{\varphi},\\ 
 &Z = \Tr e^{-\beta H} \approx e^{-\beta (E_++E_-)}. 
\end{align}
Using these, we can rewrite Eq.~\eqref{eq:rho_el_pm} as 
\begin{equation}\label{eq:rho_el_Psi_pm}
\begin{split}
 \bra{\varphi_{2j+1}&,\varphi_{2j}} \rho \ket{\varphi_{2j-1},\varphi_{2j}} \\
 \approx 
 &\biggbra{ \frac{\varphi_{2j+1}+\varphi_{2j}}{\sqrt{2}} } \Psi_+\bigg\rangle \bigg\langle \Psi_+ \biggket{ \frac{\varphi_{2j-1}+\varphi_{2j}}{\sqrt{2}} }\\
 \times 
 &\biggbra{ \frac{\varphi_{2j+1}-\varphi_{2j}}{\sqrt{2}} } \Psi_-\bigg\rangle \bigg\langle \Psi_- \biggket{ \frac{\varphi_{2j-1}-\varphi_{2j}}{\sqrt{2}} }. 
\end{split}
\end{equation}
Here, an expression of the form $\bracket{\varphi}{\Psi_\pm}$ is the representation 
of a ground state wave function in terms of the field configuration $\{\varphi(x)\}_{0\le x <L}$ along the chain, 
which we call a ``wave functional'' following Ref.~\onlinecite{Fradkin93}. 

%Equivalently, the same expression can be obtained by replacing $\rho$ by 
%$\ket{\Psi_+}\bra{\Psi_+} \otimes \ket{\Psi_-}\bra{\Psi_-}$ in Eq.~\eqref{eq:rho_el_pm}. 

%************************************************
\subsection{Calculation of reduced density matrix moments}\label{sec:rhon_wavefn_cal}
%************************************************

\newcommand{\Phit}{\tilde{\Phi}}
%----------------------------
%- Calculation of Tr rho^n
The ground state wave functional of a TLL has been derived in literature. 
\cite{Fradkin93,Stone94,Pham99,Cazalilla04,Fjaerestad08,Stephan09}
In Appendix \ref{app:wavefn_TLL}, we present its simple derivation in the operator formalism.
From Eq.~\eqref{eq:wavefn_vphi}, the wave functional is expressed as 
\begin{equation}\label{eq:wavefn_vphi_pm}
 \bracket{\varphi}{\Psi_\pm} = \frac{1}{\sqrt{\Ncal_\pm}} e^{-\frac1{K_\pm} \Ecal [\varphi]},  
\end{equation}
where $\Ecal[\varphi]$ is a quadratic functional of $\varphi$ 
[see Eq.~\eqref{eq:E_Coulomb} for the explicit form]. 
From Eq.~\eqref{eq:Ncal}, the normalization factors $\Ncal_\pm$ are given by 
\begin{equation}\label{eq:Ncal_pm}
 \Ncal_\pm = \prod_{m=1}^\infty \frac{\pi K_\pm}{k_m}. 
\end{equation}
Using Eqs.~\eqref{eq:rho_el_Psi_pm} and \eqref{eq:wavefn_vphi_pm} 
and $\Ecal[\varphi]=\Ecal[-\varphi]$, 
Eq.~\eqref{eq:rho_n} is rewritten as
\begin{equation}\label{eq:Trrhon_wavefn}
\begin{split}
 &\Tr \rho_A^n = (\Ncal_+\Ncal_-)^{-n} \int \prod_{j=1}^{2n} \Dcal\varphi_j \\
 &\times
 \exp\left[
  -\sum_{j=1}^{2n} \left( 
    \frac1{K_+} \Ecal\left[ \frac{\varphi_j+\varphi_{j+1}}{\sqrt{2}} \right] 
   +\frac1{K_-} \Ecal\left[ \frac{\varphi_j-\varphi_{j+1}}{\sqrt{2}} \right]
  \right)
 \right].
\end{split}
\end{equation}
Here each term in the argument of the exponential function 
corresponds to an edge of a sheet in Fig.~\ref{fig:Z3_pm},  
and represents the probability distribution of field configurations 
in a way analogous to the Boltzmann weight. 
As shown in Eq.~\eqref{eq:wavefn_vphi_n},  
the functional $\Ecal[\varphi]$ has a very simple form 
when written in terms of the Fourier components $\{\varphit_m\}$ of $\varphi$: 
\begin{equation}
 \Ecal [\{\varphit_m\}] = \sum_{m=1}^\infty k_m |\varphit_m|^2. 
\end{equation}
Therefore, we expand $\varphi_j$ into the Fourier components $\{\varphit_{j,m}\}$ as in Eq.~\eqref{eq:vphi_expand}, 
and rewrite Eq.~\eqref{eq:Trrhon_wavefn} as 
\begin{equation}\label{eq:Trrhon_Fourier}
\begin{split}
 \Tr~\rho_A^n = & (\Ncal_+\Ncal_-)^{-n} \int \prod_{j=1}^{2n} \prod_{m=1}^\infty (d\varphit_{j,m}d\varphit_{j,m}^*) \\
 &\times\exp \left(-\sum_{m=1}^\infty \frac{2k_m}{(K_+ K_-)^{1/2}} \Phit_m^\dagger M_n \Phit_m \right),
\end{split}
\end{equation}
where $\Phit_m=(\varphit_{1,m} , \varphit_{2,m} , \dots, \varphit_{2n,m})^t$ 
and $M_n$ is a $2n\times 2n$ matrix defined as 
\begin{gather}
 M_n := 
 \begin{pmatrix}
  A         & \frac12 B &           &           & \frac12 B\\
  \frac12 B & A         & \frac12 B &           &          \\
            & \frac12 B & A         & \ddots    &          \\
            &           & \ddots    & \ddots    & \frac12 B\\
  \frac12 B &           &           & \frac12 B & A
 \end{pmatrix},\label{eq:Mn} \\
 A:=\frac12 \left( \sqrt{\frac{K_-}{K_+}}+\sqrt{\frac{K_+}{K_-}} \right),~~
 B:=\frac12 \left( \sqrt{\frac{K_-}{K_+}}-\sqrt{\frac{K_+}{K_-}} \right).
\end{gather}
Performing the Gaussian integration and using Eq.~\eqref{eq:Ncal_pm}, 
Eq.~\eqref{eq:Trrhon_Fourier} is calculated as
\begin{equation}\label{eq:Trrhon_detM}
\begin{split}
 \Tr~\rho_A^n 
 &= (\Ncal_+\Ncal_-)^{-n} 
 \prod_{m=1}^\infty \left[ \left( \frac{\pi K_+^{1/2}K_-^{1/2}}{k_m} \right)^{2n} \frac1{\det M_n} \right],\\
 &=\prod_{m=1}^\infty \left( \det M_n \right)^{-1} 
\end{split}
\end{equation}
Since $M_n$ has the same form as the Hamiltonian of a 1D tight-binding model, 
it can be easily diagonalized and its determinant is calculated as 
\begin{equation}\label{eq:detMn}
 \det M_n = \prod_{l=0}^{2n-1} \lambda_l, ~~~
 \lambda_l:=A+B\cos\left(\frac{2\pi l}{2n}\right).
\end{equation}

%----------------------------
%- Regularization
In Eq.~\eqref{eq:Trrhon_detM}, 
we have obtained an infinite product of the form $\prod_{m=1}^\infty C^{-1}$ (with $C\ge 1$), which needs to be regularized. 
We introduce a short-distance cutoff $a_0$ of the order of the lattice spacing. 
We rewrite the product as $\prod_{m\ne 0} C^{-1/2}$.  
In this expression, $m$ runs over $L/a_0-1$ modes by considering the exclusion of the zero mode. 
Therefore the product scale as $C^{1/2} e^{-\alpha L}$~(with $\alpha=(\log C)/(2a_0)>0$). 
The prefactor $C^{1/2}$ gives a cutoff-independent (and thus universal) constant. 
[A similar technique has also been used for evaluating the fidelity in a TLL in Ref.~\onlinecite{CamposVenuti09}.]
We note that the same universal constant can also be obtained by the $\zeta$-function regularization. 
Applying this argument to Eq.~\eqref{eq:Trrhon_detM}, we arrive at
\begin{equation}\label{eq:rhon_lambda}
 \Tr ~\rho_A^n = e^{-\alpha L} (\det M_n)^{1/2},
\end{equation}
where $\alpha$ is a cutoff-dependent constant. 
Note that, although we initially assumed integer $n\ge 2$, 
the final expression \eqref{eq:rhon_lambda} also contains the case of $n=1$, where $\Tr~\rho_A=1$. 
This can be seen by setting $\alpha=0$ and 
\begin{equation}
 M_2 = 
 \begin{pmatrix}
  A & B \\
  B & A
 \end{pmatrix}.
\end{equation}
Compared to Eq.~\eqref{eq:Mn}, we have $B$ instead of $\frac12 B$ in the elements 
because the elements on the subdiagonal parts and at the upper-right/lower-left corners in Eq.~\eqref{eq:Mn} are combined. 
The determinant and the eigenvalues of $M_2$ are written in the same ways as Eq.~\eqref{eq:detMn}. 

% Then the matrix element \eqref{eq:rho_el_Psi_pm} is rewritten as 
% \begin{equation} 
% \begin{split}
%  \bra{\varphi_{2j+1},&\varphi_{2j}} \rho \ket{\varphi_{2j-1},\varphi_{2j}} 
%  \approx 
%  \frac1{\Ncal_+\Ncal_-} \\
%  \times \exp &\bigg(
%    \frac1{K_+} \Ecal\left[ \frac{\varphi_{2j+1}+\varphi_{2j}}{\sqrt{2}} \right]
%   +\frac1{K_+} \Ecal\left[ \frac{\varphi_{2j-1}+\varphi_{2j}}{\sqrt{2}} \right]\\
%   +&\frac1{K_-} \Ecal\left[ \frac{\varphi_{2j+1}-\varphi_{2j}}{\sqrt{2}} \right]
%   +\frac1{K_-} \Ecal\left[ \frac{\varphi_{2j-1}-\varphi_{2j}}{\sqrt{2}} \right]
% \bigg), 
% \end{split}
% \end{equation}

%************************************************
\subsection{Expressions of R\'enyi entropies} \label{sec:Renyi_ent_expressions}
%************************************************

%----------------------------
%- Expressions for n>=2
Equation \eqref{eq:rhon_lambda} leads to a linear scaling of $S_n$ 
as a function of the chain length $L$ as in Eq.~\eqref{eq:Sn_L}. 
The coefficient $\alpha_n$ of the linear term depends on the short-distance cutoff $a_0$ and therefore is not universal. 
The subleading constant term $\gamma_n$ for integer $n\ge 2$ is obtained as 
\begin{equation}\label{eq:gamma_n}
 \gamma_n = \frac{-1}{2(n-1)} \log (\det M_n) =\frac{-1}{2(n-1)} \sum_{l=0}^{2n-1} \log \lambda_l.
\end{equation}
We see that $\gamma_n$ is determined by the underlying field theory 
and is a function of the ratio of the two TLL parameters, $K_+/K_-$.
As an example, for $n=2$, one obtains
\begin{equation}\label{eq:gamma_2}
 \gamma_2 = -\log \left[ \frac12 \left( \sqrt{\frac{K_-}{K_+}} + \sqrt{\frac{K_+}{K_-}} \right)\right]. 
\end{equation}
In the limit of $n\to\infty$, the summation over $l$ in Eq.~\eqref{eq:gamma_n} is replaced by an integral, leading to
\begin{equation}\label{eq:gamma_inf}
 \gamma_\infty = 
 - \log \left[ \frac12 \left(\sqrt{\frac{K_-}{K_+}} + \sqrt{\frac{K_+}{K_-}} \right) \right]
 - I \left( \frac{K_--K_+}{K_-+K_+} \right) 
\end{equation}
with
\begin{equation}\label{eq:I_kappa}
 I(s) = \int_0^{2\pi} \frac{d\theta}{2\pi} \log (1+s \cos\theta) 
 = \log \left( \frac{1+\sqrt{1-s^2}}{2} \right).  
\end{equation}
Here the integral was calculated as follows. 
We differentiate $I(s)$ with respect to $s$ and integrate over $\theta$: 
\begin{equation}
 \frac{dI(s)}{ds} = \int_0^{2\pi} \frac{d\theta}{2\pi} \frac{\cos\theta}{1+s \cos\theta}
 = \frac{1}{s} - \frac{1}{s\sqrt{1-s^2}}
\end{equation}
Using $I(0)=0$ and integrating this over the interval $[0,s]$ give the final expression in Eq.~\eqref{eq:I_kappa}.

%----------------------------
%- Analytic property in n->1
In the replica procedure for calculating the von Neumann entropy, 
we compute the R\'enyi entropies $S_n$ for integer $n\ge 2$, 
take the analytic continuation to real $n\in [1,\infty]$, 
and then take the limit $n\to 1^+$. 
In Eq.~\eqref{eq:gamma_n}, we cannot find any obvious way to extend the formula of $\gamma_n$ to the case of real $n$. 
Let us focus on the expression $\gammat_n := \log (\det M_n)$ appearing in Eq.~\eqref{eq:gamma_n}. 
To gain insights on the analyticity of $\gammat_n$ as a function of $n$, 
we expand it around $K_+/K_-=1$, 
which corresponds to the limit of no inter-chain coupling. 
To this end, it is useful to introduce a parameter
\begin{equation}
 \kappa := \frac{K_- - K_+}{K_- + K_+}.  
\end{equation}
Using this, $\gammat_n$ is written as
\begin{equation}
 \gammat_n = -n \log(1-\kappa^2) 
 + \sum_{l=0}^{2n-1} \log \left[ 1+\kappa\cos\left( \frac{2\pi l}{2n} \right) \right].   
\end{equation}
Expanding around $\kappa=0$ gives
\begin{equation}\label{eq:gammat_ex}
  \gammat_n = \sum_{m=1}^\infty \frac{2n-A_{n,m}}{2m} \kappa^{2m},
\end{equation}
with 
\begin{equation}
\begin{split}
 A_{n,m} &:= \sum_{l=0}^{2n-1} \cos^{2m} \left( \frac{2\pi l}{2n} \right)\\
 &= \frac{1}{2^{2m}} \sum_{k=0}^{2m} 
 \begin{pmatrix}
  2m \\ k
 \end{pmatrix}
 \sum_{l=0}^{2n-1} e^{2\pi l(k-m)/n}. 
\end{split}
\end{equation}
In the summation over $k$, only the terms where $k-m$ is an integer multiple of $n$ contribute. 
For $m<n$, it occurs only for $k=m$, 
and  $A_{n,m}$ is given by a simple expression
\begin{equation}
 A_{n,m} = \frac{2n}{2^{2m}}  
 \begin{pmatrix}
  2m \\ m
 \end{pmatrix}.
\end{equation}
For $m\ge n$, $A_{n,m}$ can contain other terms 
and show nontrivial dependences on $n$ and $m$. 
For example, for $m=1$, one obtains
\begin{equation}
 A_{n,1} = 
 \begin{cases}
  2 & (n=1)    \\
  n & (n\ge 2) .
 \end{cases}
\end{equation}
This leads to the lowest-order expansion of $\gammat_n$ for $n\ge 2$
\begin{equation}\label{eq:gammat_ex_2}
 \gammat_n = \frac{n}2 \kappa^2 + {\cal O}(\kappa^4), 
\end{equation}
which is {\it not} smoothly connected to $\gammat_1=0$ as $n\to 1^+$. 
Multiplying $-1/[2(n-1)]$ to Eq.~\eqref{eq:gammat_ex_2}, we obtain
\begin{equation}\label{eq:gamman_kappa2}
 \gamma_n = - \frac{n}{4(n-1)} \kappa^2 + {\cal O}(\kappa^4 )
\end{equation}
for $n\ge 2$. 
If we naively take the limit $n\to 1^+$ in this expression, 
we find that the coefficient of the leading (order-$\kappa^2$) term in $\gamma_n$ is divergent. 
This indicates that in this problem, it is not easy to study the von Neumann entropy $S_1$ 
from the knowledge of the R\'enyi entropies $S_n$ with integer $n\ge 2$. 
At present, we do not have any analytic prediction on $S_1$.  
However, our numerical result in Sec.~\ref{sec:numerics} indicates 
that $S_1$ also obeys a linear function of $L$ 
and that the subleading constant $\gamma_1$ is determined by $K_+/K_-$. 
In particular, for small $\kappa$, we find that $\gamma_1$ obeys a non-trivial power function 
\begin{equation}\label{eq:gamma1_kappa}
 \gamma_1 \approx - a\kappa^b
\end{equation}
with $b\approx 1.6$-$1.7$.
In spite of the qualitative difference between Eq.~\eqref{eq:gamman_kappa2} and Eq.~\eqref{eq:gamma1_kappa}, 
our numerical result also suggests that for fixed $\kappa$,
$\gamma_n$ changes rather smoothly when $n$ is changed from $2$ to $1$. 
In Sec.~\ref{sec:numerics}, we will present some possible scenarios 
as to  how these two different small-$\kappa$ behaviors are connected to each other.

The current problem adds to the list of problems 
where the R\'enyi entropy shows quite a non-trivial analyticity as a function of $n$. 
Here we cite a few examples known in literature. 
In massive integrable quantum field theory,\cite{Cardy07}  
the analytic continuation could not be uniquely introduced from the knowledge for integer $n\ge 2$, 
and the appropriate one needed to be chosen carefully. 
In the R\'enyi entanglement entropy of two disjoint intervals in CFT,\cite{Calabrese10} 
the analytic form for integer $n\ge 2$ has a non-trivial form, 
and its analytic continuation to $n\to 1^+$ has been achieved in certain limits, 
leaving a general solution open. 
In the R\'enyi entropy of a line embedded in 2D Ising models,\cite{Stephan10} 
the constant part behaves as a step-like function of $n$ with discontinuity at $n=1$, 
which means that the standard replica procedure fails to address the case of $n=1$.

%************************************************
\section{Boundary conformal field theory approach} \label{sec:BCFT}
%************************************************

%----------------------------
%- Section introduction
In this section we express the partition functions, $Z_n$ and $Z^n$, 
in Eq.~\eqref{eq:rho_n_Zn} as the transition amplitudes between conformal boundary states. 
In the limit $\beta\gg L\gg 1$, these partition functions contain universal multiplicative constant contributions,   
known as the boundary ``ground-state degeneracies.''\cite{Affleck91}
This approach does not require any regularization procedure 
and determines the universal contributions in the partition functions 
in a way consistent with a certain condition under the modular transformation 
(Cardy's consistency condition\cite{Cardy89}). 
Similar approaches were also used quite recently 
to calculate the entanglement entropy in 2D critical wave functions\cite{Hsu10,Oshikawa10}
and the ground-state fidelity in TLLs.\cite{CamposVenuti09} 

%############################
\begin{figure*}
\includegraphics[width=0.9\textwidth]{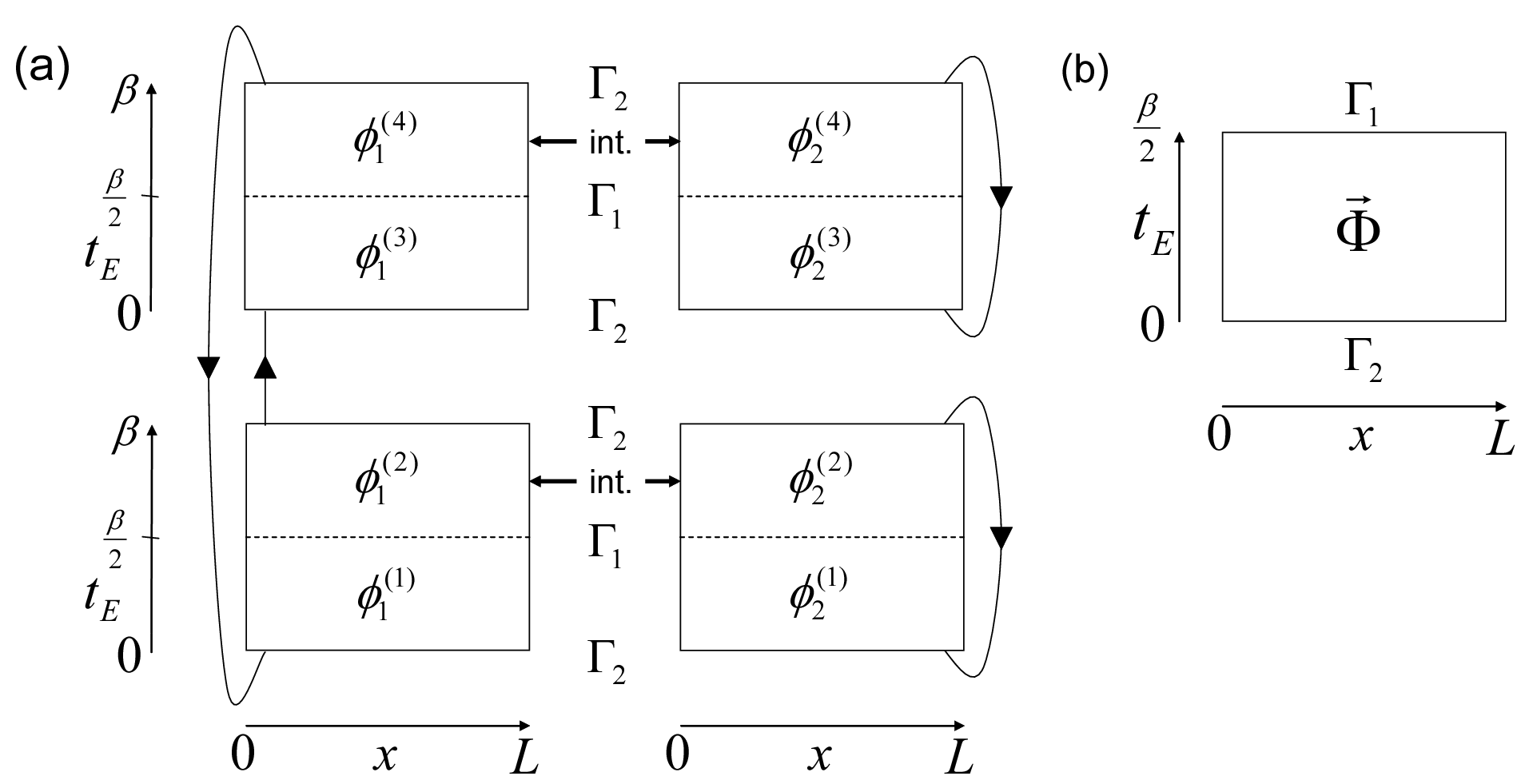}%{Z2_g1g2.eps}
%\quad
%\includegraphics[width=0.3\textwidth]{Z2_Phi.eps}
\caption{Two representations of $Z_2$.
Four sheets are interconnected in (a). 
We fold each sheet at $t_E=\beta/2$ and superpose all the 8 pieces of sheets, 
leading to a 8-component field $\Phiv$ on a single sheet in (b). }
\label{fig:Z2}
\end{figure*}
%############################

%************************************************
\subsection{Compactification conditions of bosonic fields}
%************************************************

%----------------------------
%- Subsection introduction
To apply boundary CFT to the system introduced in Sec.~\ref{sec:CTLL}, 
one needs to precisely discuss the compactification conditions imposed on the bosonic fields. 
We will see that zero modes of the symmetric/antisymmetric channels are intertwined and require a careful treatment. 

%----------------------------
%- Compactification conditions on the original fields
The original bosonic fields, $\phi_\chlab$ and $\theta_\chlab$ ($\chlab=1,2$),
defined along the chains are compactified on circles with different radii. 
When periodic boundary conditions (PBC) are imposed, 
these fields can acquire winding numbers when going around the chains, namely,\cite{comment_compactification} 
\begin{equation} \label{eq:comp_ori}
\begin{split}
 & \phi_\chlab (L)  = \phi_\chlab (0)   + 2\pi r n_\chlab,\\
 & \theta_\chlab (L) = \theta_\chlab (0) + 2\pi \tilde{r} m_\chlab,\\
 & n_\chlab,m_\chlab \in \mathbb{Z}.
\end{split} 
\end{equation}
Here the compactification radii are given by 
\begin{equation}
 r=\frac1{2\sqrt{\pi}}, \quad \tilde{r}=\frac{1}{\sqrt{\pi}}. 
\end{equation}

%----------------------------
%- Recovering of conformal invariance
Before discussing the compactification conditions in the symmetric/antisymmetric channels, 
let us mention that 
$H$ in Eq.~\eqref{eq:H+H-} is not a Hamiltonian of a conformally invariant system 
because the two velocities $v_\pm$ in Eq.~\eqref{eq:H+-} are different in general. 
To apply boundary CFT later, we restore the conformal invariance by simply replacing $v_\pm\to 1$. 
Although this replacement changes the spectrum of the Hamiltonian, 
it does not change the eigenstates. 
The ground state and therefore its entanglement properties should remain unchanged. 
To further simplify the Hamiltonian, we rescale the bosonic fields as 
$\Phi_\pm = \phi_\pm/\sqrt{K_\pm}$ and  
$\Theta_\pm = \sqrt{K_\pm}\theta_\pm$. 
The new Hamiltonian is 
\begin{equation}\label{eq:Htilde}
 \tilde{H} = \frac12 \int_0^L dx ~\left[
 \left(\frac{d\vec{\Theta}}{dx}\right)^2 + \left(\frac{d\vec{\Phi}}{dx}\right)^2
 \right]
\end{equation}
with 
\begin{equation}
 \vec{\Phi} = \begin{pmatrix} \Phi_+ \\ \Phi_- \end{pmatrix} ,~~
 \vec{\Theta} = \begin{pmatrix} \Theta_+ \\ \Theta_- \end{pmatrix}. 
\end{equation}

%----------------------------
%- Compactification conditions on the new fields
From Eq.~\eqref{eq:comp_ori}, the new bosonic fields are subject to the conditions 
\begin{align}
 &\vec{\Phi}(L) = \vec{\Phi}(0)+2\pi \uvec,  \quad \uvec=n_1 \vec{a}_1 + n_2 \vec{a}_2, \label{eq:comp_u} \\ 
 &\vec{\Theta}(L) =  \vec{\Theta}(0)+2\pi \vvec,  \quad \vvec=m_1 \vec{b}_1 + m_2 \vec{b}_2, \label{eq:comp_v} \\
 &n_1,n_2,m_1,m_2 \in \mathbb{Z},\notag  
\end{align}
where 
\begin{equation}
 \vec{a}_{1,2} = \frac{r}{\sqrt{2}} 
 \begin{pmatrix}
  1/\sqrt{K_+} \\ \pm 1/\sqrt{K_-}
 \end{pmatrix}, \quad
 \vec{b}_{1,2} = \frac{\tilde{r}}{\sqrt{2}} 
 \begin{pmatrix}
  \sqrt{K_+} \\ \pm \sqrt{K_-}. 
 \end{pmatrix}
\end{equation}
Note that 
$\vec{a}_i\cdot \vec{b}_j=\frac{1}{2\pi}\delta_{ij}$. 
Let $\Lambda$ be the lattice of $\vec{u}$ defined by Eq.~\eqref{eq:comp_u}, 
and let $\Lambda^*$ be its reciprocal lattice. 
Then $\vec{v}$ defined by Eq.~\eqref{eq:comp_v} lives on $\frac{1}{2\pi} \Lambda^*$.
The lattices of $\uvec$ and $\vvec$ introduced in this way are called the compactification lattices of $\Phiv$ and $\Thetav$.

%************************************************
\subsection{Reduced density matrix moments and boundary states}  \label{sec:entropy_boundary_state}
%************************************************

%----------------------------
%- Sheet folding -> 8-component boson
We consider the partition functions, $Z_n$ and $Z^n$, appearing in Eq.~\eqref{eq:rho_n_Zn}. 
In the following discussions, we mainly focus on the case $n=2$; 
the generalization to arbitrary integer $n\ge 1$ is straightforward 
and will be done in Sec.~\ref{sec:Gamma12}. 
From the argument in Sec.~\ref{sec:path_rdm}, 
$Z_2$ is expressed by four sheets interconnected as shown in Fig.~\ref{fig:Z2}(a). 
Invoking the folding technique of Refs.~\onlinecite{Oshikawa10,Oshikawa97}, 
we fold each sheet at $t_E=\beta/2$ and superpose all the 8 pieces of sheets. 
As a result, we have a 8-component bosonic field $\Phiv$ living 
on a cylinder of lengths $L$ and $\beta/2$ 
in the spatial and temporal directions respectively; see Fig.~\ref{fig:Z2}(b). 
The Hamiltonian $\Ht$ for this ``system'' is written in the same form as in Eq.~\eqref{eq:Htilde}, 
but now $\Phiv$ and $\Thetav$ consist of 8 components each:
\begin{align}
 &\Phiv = (\Phi_+^{(1)},\Phi_-^{(1)},\Phi_+^{(2)},\Phi_-^{(2)},\Phi_+^{(3)},\Phi_-^{(3)},\Phi_+^{(4)},\Phi_-^{(4)})^t, \\
 &\Thetav = (\Theta_+^{(1)},\Theta_-^{(1)},\Theta_+^{(2)},\Theta_-^{(2)},\Theta_+^{(3)},\Theta_-^{(3)},\Theta_+^{(4)},\Theta_-^{(4)})^t. 
\end{align} 
Here, the components of $\Phiv$ are related to $\phi_\chlab^{(j)}$'s in Fig.~\ref{fig:Z2}(a) as
\begin{equation}
 \Phi_\pm^{(j)} = \frac1{\sqrt{2K_\pm}} (\phi_1^{(j)} \pm \phi_2^{(j)}),  
\end{equation} 
and $\Theta_\pm^{(j)}$ are defined as their dual counterparts. 
The 8-component fields are subject to the conditions 
\begin{align}
 &\Phiv(L) = \Phiv(0) +2\pi \uvec, \quad \uvec\in\Xi\equiv \Lambda^4\\
 &\Thetav(L) = \Thetav(0) +2\pi \vvec, \quad \vvec\in\Xit\equiv \left( \frac1{2\pi} \Lambda^* \right)^4. 
\end{align} 
The primitive vectors of the lattice $\Xi$ are $\avec_\chlab^{(j)} ~(\chlab=1,2;~j=1,2,3,4)$, 
each of which is defined by inserting $\avec_\chlab$ into $(2j-1)$- and $(2j)$-th elements and zeros into the others. 
Similarly, the primitive vectors of the lattice $\Xit$ are $\bvec_\chlab^{(j)}~(\chlab=1,2;~j=1,2,3,4)$, defined likewise from $\bvec_\chlab$. 

%----------------------------
%- Boundary conditions
At the two boundaries at $t_E=\beta/2$ and $0$, 
the following boundary conditions are imposed respectively: 
\begin{align}
 \Gamma_1:~& \phi_1^{(2j-1)}=\phi_1^{(2j)}, ~\phi_2^{(2j-1)}=\phi_2^{(2j)}~~(j=1,2), \label{eq:Gamma1}\\
 \Gamma_2:~& \phi_1^{(2j)}=\phi_1^{(2j+1)}, ~\phi_2^{(2j-1)}=\phi_2^{(2j)} \notag \\
           & (j=1,2;~ \phi_1^{(5)}\equiv\phi_1^{(1)}). \label{eq:Gamma2}
\end{align}
We will express these conditions using boundary states, $\ket{\Gamma_1}$ and $\ket{\Gamma_2}$. 
The partition functions we wish to calculate are expressed as the transition amplitudes between these states:  
\begin{align}
 Z_2 = Z_{\Gamma_1 \Gamma_2} = \bra{\Gamma_1} e^{-\frac{\beta}2 \Ht} \ket{\Gamma_2},\\
 Z^2 = Z_{\Gamma_1 \Gamma_1} = \bra{\Gamma_1} e^{-\frac{\beta}2 \Ht} \ket{\Gamma_1}.  
\end{align}

%************************************************
\subsection{Boundary state formalism}\label{sec:boundary_state}
%************************************************

%----------------------------
%- Boundary CFT for multicomponent boson
Before considering the two boundary states $\ket{\Gamma_{1,2}}$ in more detail, 
we discuss the construction of boundary states in a more general setting. 
The boundary CFT for multicomponent bosons 
has been developed in string theory\cite{Polchinski88,Callan88,Ooguri96} 
and applied to condensed matter problems.\cite{Wong94,Affleck01,Chamon03,Oshikawa06}
In particular, a ``mixed'' Dirichlet/Neumann boundary condition, which we focus on here, 
has been discussed in Refs.~\onlinecite{Ooguri96,Oshikawa06}. 
Such ``mixed'' conditions have recently been applied 
to the calculation of the entanglement entropy in 2D critical wave functions.\cite{Oshikawa10} 
%Our calculation is very close in spirit to that of Ref.~\onlinecite{Oshikawa10}. 
Here we review basic knowledge on the boundary CFT for multicomponent bosons, 
and discuss how to construct the boundary state for a ``mixed'' Dirichlet/Neumann condition. 
For further details, we refer the reader to, e.g., Refs.~\onlinecite{Affleck01,Oshikawa06,Oshikawa10} 
(especially Ref.~\onlinecite{Oshikawa10} for the present application\cite{notation_boson}), 
which contain useful summaries of boundary CFT for multicomponent bosons.
The main result of this subsection is the formula of the boundary ``ground-state degeneracy'' in Eq.~\eqref{eq:g_Gamma}, 
which is used later to calculate universal (non-extensive) constant contributions in partition functions.

%----------------------------
%- Mode expansions of bosonic fields
We consider a $c$-component free boson 
defined by the Hamiltonian $\Ht$ in Eq.~\eqref{eq:Htilde}. 
The system is placed on a cylinder like Fig.~\ref{fig:Z2}(b) 
and we impose certain conformally invariant boundary conditions at both ends. 
Since the PBC is imposed in the $x$ direction, 
the bosonic fields have the following mode expansions:
\begin{subequations}\label{eq:PhivThetav_ex}
\begin{align}
 &\Phiv(t,x) = \Phiv_0 + \frac{2\pi}L (\uvech x+ \vvech t) \\
 &~~+ \sum_{m=1}^\infty \frac1{\sqrt{4\pi m}} \left( \avec_m^L e^{-ik_m(x+t)} + \avec_m^R e^{ik_m(x-t)}+{\rm h.c.} \right), \notag\\
 &\Thetav(t,x) = \Thetav_0 + \frac{2\pi}L (\vvech x+ \uvech t) \\
 &~~+ \sum_{m=1}^\infty \frac1{\sqrt{4\pi m}} \left( \avec_m^L e^{-ik_m(x+t)} - \avec_m^R e^{ik_m(x-t)}+{\rm h.c.} \right), \notag
\end{align}
\end{subequations}
with $k_m=2\pi m/L$. 
The spectra of $\uvech$ and $\vvech$ 
form the lattices $\Xi$ and $\Xit = \frac{1}{2\pi} \Xi^*$, respectively. 
We have included the dependence on the real time $t$, 
which help to see that $\avec_m^L$ ($\avec_m^R$) represents a left (right) moving mode. 
The elements of vectors, which we label by $j=1,2,\dots,c$, obey the commutation relations
\begin{align}
 &[\Phi_{0,j},\hat{v}_{j'}]=[\Theta_{0,j},\hat{u}_{j'}]=i\delta_{jj'}/(2\pi),\\
 &[a_{m,j}^L,a_{m',j'}^{L\dagger}]=[a_{m,j}^R,a_{m',j'}^{R\dagger}]=\delta_{mm'}\delta_{jj'}.
\end{align}
Using the expansions \eqref{eq:PhivThetav_ex}, 
the Hamiltonian $\Ht$ is diagonalized as
\begin{equation}
 \Ht = \frac{2\pi}L \bigg[ 
  \pi (\uvech^2 + \vvech^2) 
  + \sum_{m=1}^\infty m \left( \avec_m^{L\dagger}\cdot\avec_m^L + \avec_m^{R\dagger}\cdot\avec_m^R \right)
  - \frac{c}{12}
 \bigg],
\end{equation}
where the last term comes from the zero-point motions of oscillators (Casimir effect). 
The ground state $\ket{\Psi}$ of $\Ht$ is given by the condition 
$\avec_m^{L/R} \ket{\Psi}=\uvech\ket{\Psi}=\vvech\ket{\Psi}=\vec{0}$. 
We can decompose Eq.~\eqref{eq:PhivThetav_ex} into the chiral components as
\begin{equation}
\begin{split}
 \Phiv(t,x)&= \Phiv_L(x_+) + \Phiv_R(x_-),\\
 \Thetav(t,x)&= \Phiv_L(x_+) - \Phiv_R(x_-),\\
 x_\pm &= t\pm x.
\end{split}
\end{equation}
with 
\begin{equation}\label{eq:PhiLR_ex}
\begin{split}
 \Phiv_{L/R}(x_\pm)&=\frac12 (\Phiv_0 \pm \Thetav_0) + \frac{\pi}L \left( \pm \uvech + \vvech \right) x_\pm \\
 &+ \sum_{m=1}^\infty \frac1{\sqrt{4\pi m}} \left( \avec_m^{L/R} e^{-ik_m x_\pm} +{\rm h.c.} \right).
\end{split}
\end{equation}

%----------------------------
%- Boundary state 
We now introduce a conformally invariant boundary condition $\Gamma$ at the time $t=0$. 
Boundary conformal invariance implies that the momentum density operator $T_L-T_R$ vanishes at the boundary. 
Here, $T_{L/R}(t,x)=T_{L/R}(x_\pm) =2\pi (\partial_\pm \Phiv)^2$ (with $\partial_\pm := \partial_{x_\pm}$)
are the chiral components of the energy-momentum tensor. 
The conformal boundary state $\ket{\Gamma}$ therefore satisfies
\begin{equation}\label{eq:Gamma_T}
 \left[ T_L(x) - T_R(x) \right] \ket{\Gamma} = 0 .
\end{equation}
[Here, $T_{L/R}(x)$ is defined by $T_{L/R}(t=0,x)$. The same convention applies to $\Phi_{L/R}(x)$ and $J_{L/R}(x)$ below.] 
In a multicomponent boson, one can also introduce additional symmetry requirement of the form
\begin{equation}\label{eq:Gamma_J}
 \left[ \vec{J}_L(x)- \Rcal \vec{J}_R(x) \right] \ket{\Gamma} = 0, 
\end{equation}
which represents the conservation of currents in a general form 
(associated with a Heisenberg algebra). 
Here $\Rcal$ is an orthogonal matrix, and 
\begin{equation}
 \vec{J}_{L/R} (t,x) = \vec{J}_{L/R}(x_\pm) = \pm \partial_\pm \Thetav(t,x) = \partial_\pm \Phiv_{L/R}(x_\pm)
\end{equation}
are the chiral components of the current operator. 
Since $T_{L/R}(x_\pm) = 2\pi (J_{L/R}(x_\pm))^2$, 
Eq.~\eqref{eq:Gamma_J} implies Eq.~\eqref{eq:Gamma_T}. 
Therefore, Eq.~\eqref{eq:Gamma_J} defines a subclass of conformal boundary states for multicomponent bosons, 
which have many interesting physical applications.\cite{Wong94,Chamon03,Oshikawa06} 
On the other hand, conformal boundary states which satisfy only Eq.~\eqref{eq:Gamma_T} and not Eq.~\eqref{eq:Gamma_J} 
are also known.\cite{Affleck01}

%----------------------------
%- Current-conserving subclass
We now focus on the subclass defined by Eq.~\eqref{eq:Gamma_J}. 
The condition can be rewritten as
\begin{equation}\label{eq:Gamma_Phi}
 \partial_x \left[ \Phiv_L(x) + \Rcal \Phiv_R(x) \right] \ket{\Gamma} = 0.
\end{equation}
This means that $\Phiv_L + \Rcal \Phiv_R$ is fixed at a constant vector along the boundary. 
In particular, setting $\Rcal$ to the identity matrix $I$, 
we have the Dirichlet boundary condition (``D''), 
where $\Phiv$ is fixed at a constant vector along the boundary. 
Setting $\Rcal=-I$ leads to fixing $\Thetav$, 
which then means the Neumann boundary condition (``N'') $\partial_t \Phiv =0$
(since $\partial_t \Phiv = \partial_x \Thetav$). 

%----------------------------
%- How to construct |Gamma>
To obtain the explicit form of $\ket{\Gamma}$, 
we decompose Eq.~\eqref{eq:Gamma_Phi} into Fourier components 
using Eq.~\eqref{eq:PhiLR_ex}, leading to
\begin{subequations}
\begin{gather}
 \left[ (\uvech + \vvech) + \Rcal (\uvech-\vvech) \right] \ket{\Gamma} = 0, \label{eq:Gamma_def_1} \\
  (\avec_m^L+ \Rcal \avec_m^{R\dagger})\ket{\Gamma}
 =(\avec_m^{L\dagger}+ \Rcal \avec_m^R)\ket{\Gamma} = 0. \label{eq:Gamma_def_2}
\end{gather}
\end{subequations}
The solution of Eq.~\eqref{eq:Gamma_def_2} is given by the Ishibashi state\cite{Ishibashi89}
\begin{equation} \label{eq:Ishibashi}
 |(\uvec,\vvec)\rangle\rangle 
 := \exp\left( 
 -\sum_{m=1}^\infty \avec_m^{L\dagger} \cdot \Rcal \avec_m^{R\dagger} 
 \right)
 \ket{(\uvec,\vvec)}, 
\end{equation}
where $\ket{(\uvec,\vvec)}$ is an oscillator vacuum 
characterized by the zero mode quantum numbers (or ``winding numbers'')
$\uvec\in\Xi$ and $\vvec\in\Xit$. 
If $(\uvec,\vvec)$ satisfies 
\begin{equation}\label{eq:condition_uv}
 (\uvec + \vvec) + \Rcal (\uvec-\vvec) = 0 
\end{equation}
required from Eq.~\eqref{eq:Gamma_def_1}, 
the Ishibashi state $|(\uvec,\vvec)\rangle\rangle$ satisfies the conformal invariance. 
It is known, however, that in order to obtain a stable boundary state for a given $\Rcal$, 
one must take a linear combination of the Ishibashi states 
over all possible $(\uvec,\vvec)$ satisfying Eq.~\eqref{eq:condition_uv}.\cite{Oshikawa06} 

%----------------------------
%- D-N mixed boundary condition
We proceed our discussion focusing on the case of a ``mixed'' Dirichlet/Neumann boundary condition, 
which is defined as a special case of Eq.~\eqref{eq:Gamma_Phi} as follows.  
In the $c$-dimensional space of the vectorial bosonic fields, 
we impose ``D'' for the $d_D$-dimensional subspace $V_D$ 
and ``N'' for the remaining $d_N(=c-d_D)$-dimensional subspace $V_N$ perpendicular to it. 
Namely, 
\begin{align}
 &\svec \cdot \partial_x \Phiv  (x) \ket{\Gamma} = 0~~\text{for}~\svec\in V_D,\\
 &\svec \cdot \partial_x \Thetav(x) \ket{\Gamma} = 0~~\text{for}~\svec\in V_N. 
\end{align}
Let $P_\parallel$ and $P_\perp$ be the projection operators onto $V_D$ and $V_N$, respectively. 
Then $\Rcal$ in Eq.~\eqref{eq:Gamma_Phi} is expressed as
\begin{equation}\label{eq:Rcal_def}
 \Rcal = I P_\parallel + (-I) P_\perp = P_\parallel - P_\perp, 
\end{equation}
which is the reflection operator about the ``surface'' $V_D$. 
As explained above, the corresponding boundary state $\ket{\Gamma}$ is constructed 
as a linear combination of Ishibashi states \eqref{eq:Ishibashi}:
\begin{equation}\label{eq:Gamma_general}
 \ket{\Gamma} = g_\Gamma \sum_{(\uvec,\vvec)} |(\uvec,\vvec)\rangle\rangle,
\end{equation}
where $g_\Gamma$ is a prefactor to be determined later 
and the summation runs over all possible $(\uvec,\vvec)$ satisfying Eq.~\eqref{eq:condition_uv}. 
Usually, instead of the condition \eqref{eq:condition_uv}, 
it is sufficient to require separate conditions for $\uvec$ and $\vvec$: 
\begin{equation}\label{eq:condition_u_v}
 \Rcal \uvec = -\uvec, \quad
 \Rcal \vvec = \vvec. 
\end{equation}
Since $\uvec$ and $\vvec$ live on different lattices $\Xi$ and $\Xit$, 
a solution $(\uvec,\vvec)$ satisfying only Eq.~\eqref{eq:condition_uv} and not Eq.~\eqref{eq:condition_u_v} 
appears only when the primitive vectors of the lattices are fine-tuned, 
and is not considered in the present discussion. 
Because of the definition \eqref{eq:Rcal_def} of $\Rcal$ in the present case, 
the conditions \eqref{eq:condition_u_v} imply 
\begin{equation}
 \uvec\in V_N, \quad \vvec\in V_D. 
\end{equation}
Let $\Xi_N$ be the set of $\uvec\in \Xi$ satisfying $\uvec\in V_N$ 
and $\Xit_D$ be the set of $\vvec\in \Xit$ satisfying $\vvec\in V_D$. 
Then, Eq.~\eqref{eq:Gamma_general} is rewritten as 
\begin{equation}\label{eq:Gamma_state}
 \ket{\Gamma} = g_\Gamma \sum_{\uvec\in \Xi_N} \sum_{\vvec\in \Xit_D} |(\uvec,\vvec)\rangle\rangle. 
\end{equation}

%----------------------------
%- Cardiy's consistency condition
The prefactor $g_\Gamma$ is fixed by requiring Cardy's consistency condition,\cite{Cardy89} stated as follows. 
We impose boundary conditions $\Gamma$ and $\Gamma'$ at the imaginary time $t_E=\beta/2$ and $0$ respectively, 
and consider the transition amplitude (partition function):
\begin{equation}
 Z_{\Gamma\Gamma'} = \bra{\Gamma} e^{-\frac{\beta}2 \Ht} \ket{\Gamma'}.
\end{equation}
This can be expressed as a function of 
\begin{equation}
 q=e^{2\pi i \tau} = e^{-2\pi\beta/L}, 
\end{equation}
where $\tau=i\beta/L$ is the modular parameter\cite{modular_para} 
(this picture is referred to as the ``closed string channel''). 
By modular transformation, we exchange the roles of space and time 
and express $Z_{\Gamma\Gamma'}$ as a function of 
$\qt=e^{-2\pi i/\tau}=e^{-2\pi L/\beta}$ (``open string channel''). 
In this picture, we may define the Hamiltonian $\Ht_{\Gamma\Gamma'}$ 
for a 1D system with two boundary conditions $\Gamma$ and $\Gamma'$ at the ends, 
and write the partition function as 
$Z_{\Gamma\Gamma'}(\qt)=\Tr~e^{-L \Ht_{\Gamma\Gamma'}}$. 
This means that $Z_{\Gamma\Gamma'}(\qt)$ is determined by the spectrum of $\Ht_{\Gamma\Gamma'}$. 
Therefore it should have the form
\begin{equation}
 Z_{\Gamma\Gamma'} (\qt) = \sum_h N_{\Gamma\Gamma'}^h \chi_h^{\rm Vir} (\qt),
\end{equation}
where $\chi_h^{\rm Vir} (\qt)$ is a character of the Virasoro algebra. 
The coefficient $N_{\Gamma\Gamma'}^h$ can be interpreted as the number of primary fields with conformal weight $h$, 
and has to be a non-negative integer (Cardy's condition\cite{Cardy89}). 
Usually it is also required that $N_{\Gamma\Gamma}^0=1$, where $h=0$ corresponds to the identity operator. 
This is related to the uniqueness of the ground state of $\Ht_{\Gamma\Gamma'}$. 
This requirement can be used to fix $g_\Gamma$. 

%----------------------------
%- Gamma-Gamma amplitude
Now we calculate the amplitude between two $\ket{\Gamma}$'s defined by Eq.~\eqref{eq:Gamma_state}:
\begin{equation}
\begin{split}
 Z_{\Gamma\Gamma} (q) 
  = g_\Gamma^2 \left( \frac1{\eta(q)} \right)^c \sum_{\uvec\in \Xi_N} \sum_{\vvec\in \Xit_D} q^{\frac\pi2 (\uvec^2+\vvec^2)}, 
\end{split}
\end{equation}
where
\begin{equation}
 \eta (q) = q^{1/24} \prod_{m=1}^\infty (1-q^m)
\end{equation}
is the Dedekind $\eta$ function. 
By modular transformation, 
we can rewrite $Z_{\Gamma\Gamma}$ using $\qt$:
\begin{equation}
\begin{split}
 Z_{\Gamma\Gamma} (\qt)
 &= g_\Gamma^2 \pi^{-c/2} v_0(\Xi_N)^{-1} v_0(\Xit_D)^{-1} \\
 &\times \left( \frac1{\eta(\qt)} \right)^c
 \sum_{\rvec\in \Xi_N^*} \sum_{\svec\in \Xit_D^*} \qt^{\frac1{2\pi}(\rvec^2+\svec^2)},  
\end{split}
\end{equation} 
where $v_0(...)$ represents the unit cell volume of the lattice. 
Here we have used the following identities:
\begin{gather}
 \eta(q) = \left( \frac{\beta}L \right)^{1/2} \eta (\qt), \\
 \sum_{\uvec\in\Xi_N} q^{\frac{\pi}2 \uvec^2} 
 = \frac1{v_0(\Xi_N)} \left( \frac{\beta}{\pi L} \right)^{d_N/2} 
   \sum_{\rvec\in\Xi_N^*} \qt^{\frac1{2\pi} \rvec^2},\\
 \sum_{\vvec\in\Xit_D} q^{\frac{\pi}2 \vvec^2} 
 = \frac1{v_0(\Xit_D)} \left( \frac{\beta}{\pi L} \right)^{d_D/2} 
   \sum_{\svec\in\Xit_D^*} \qt^{\frac1{2\pi} \svec^2}.
\end{gather}
The second and third equations come from the multi-dimensional generalization of the Poisson summation formula. 
To satisfy Cardy's consistency condition above, 
we require the coefficient of the term with $(\rvec,\svec)=(\vec{0},\vec{0})$ to be unity, obtaining 
\begin{equation}\label{eq:g_Gamma}
 g_\Gamma = \pi^{c/4} v_0(\Xi_N)^{1/2} v_0(\Xit_D)^{1/2}. 
\end{equation}

%----------------------------
%- Boundary ``ground-state degeneracy''
The constant $g_\Gamma$ appears in the overlap between the ground state $\ket{\Psi}$ of $\Ht$ and the boundary state \eqref{eq:Gamma_state}:
$g_\Gamma = \bracket{\Psi}{\Gamma}$. 
This means that the partition function $Z_{\Gamma\Gamma'}$ has 
a multiplicative constant contribution $g_\Gamma g_{\Gamma'}$ coming from the boundaries 
in the limit $\beta/2 \gg L \gg 1$. 
This result can be interpreted as follows. 
In the open string channel picture, the 1D system described by $\Ht_{\Gamma\Gamma'}$ 
has the ``spacial length'' $\beta/2$ and the ``inverse temperature'' $L$. 
The ground state of $\Ht_{\Gamma\Gamma'}$ is unique, and therefore the thermal entropy goes to zero in the ``zero temperature'' limit $1/L\to 0$. 
On the other hand, when $\beta/2 \gg L \gg 1$, the ``temperature'' $1/L$ is high enough 
and the spectrum of $\Ht_{\Gamma\Gamma'}$ looks effectively continuous. 
In this case, the thermal entropy acquires a constant contribution $\log (g_\Gamma g_{\Gamma'})$, 
in addition to the standard extensive contribution linear in temperature.
Because of this, $g_\Gamma$ is referred to as the boundary ``ground-state degeneracy,'' 
and is generally non-integer.\cite{Affleck91}

%************************************************
\subsection{Boundary conditions $\Gamma_1$ and $\Gamma_2$} \label{sec:Gamma12}
%************************************************

%----------------------------
%- Condition Gamma2
The two boundary conditions $\Gamma_{1,2}$ in Eqs.~\eqref{eq:Gamma1} and \eqref{eq:Gamma2}  
can be expressed as special cases of ``mixed'' Dirichlet/Neumann conditions discussed in the previous subsection. 
The condition $\Gamma_2$ in Eq.~\eqref{eq:Gamma2} leads to the following ``D'' conditions:
\begin{equation}
\begin{split}
 & 0=\partial_x (\phi_1^{(2j)}-\phi_1^{(2j+1)}) = \frac1{\rt} (\bvec_1^{(2j)}-\bvec_1^{(2j+1)}) \cdot \partial_x \Phiv, \\
 & 0=\partial_x (\phi_2^{(2j-1)}-\phi_2^{(2j)}) = \frac1{\rt} (\bvec_2^{(2j-1)}-\bvec_2^{(2j)}) \cdot \partial_x \Phiv  \\ 
 & (j=1,2;~ \bvec_1^{(5)}\equiv\bvec_1^{(1)}) .
\end{split}
\end{equation}
Therefore, the subspace $V_{D2}$ where ``D'' is imposed is spanned by the following (non-orthogonal) basis vectors:
\begin{equation}\label{eq:svec}
\begin{split}
 \svec_1 = \bvec_2^{(1)}-\bvec_2^{(2)}, \quad 
 \svec_2 = \bvec_1^{(2)}-\bvec_1^{(3)}, \\
 \svec_3 = \bvec_2^{(3)}-\bvec_2^{(4)}, \quad
 \svec_4 = \bvec_1^{(4)}-\bvec_1^{(1)}.
\end{split}
\end{equation}
In the perpendicular space $V_{N2}$, 
the field $\Phiv$ is free at the boundary, 
and therefore we impose ``N'', where the dual field $\Thetav$ is locked. 
This space is spanned by 
\begin{equation}\label{eq:tvec}
\begin{split} 
 \tvec_1 = \avec_2^{(1)}+\avec_2^{(2)}, \quad 
 \tvec_2 = \avec_1^{(2)}+\avec_1^{(3)}, \\
 \tvec_3 = \avec_2^{(3)}+\avec_2^{(4)}, \quad
 \tvec_4 = \avec_1^{(4)}+\avec_1^{(1)}.
\end{split}
\end{equation}
Now we consider the lattices $\Xi_{N2}$ and $\Xit_{D2}$ used to construct 
the boundary state $\ket{\Gamma_2}$ as in Eq.~\eqref{eq:Gamma_state}. 
Since any linear combination of $\{\svec_j\}$ with integer coefficients 
belongs to the lattice $\Xit$, 
$\{\svec_j\}$ can be used as the primitive vectors of $\Xit_{D2} ~(=\Xit \cap V_{D2})$. 
Similarly, $\{\tvec_j \}$ can be used as the primitive vectors of $\Xi_{N2} ~(= \Xi \cap V_{N2})$. 

%----------------------------
%- Winding numbers
What are the meanings of the lattices $\Xi_{N2}$ and $\Xit_{D2}$ introduced in this way?
Initially, in the mode expansions \eqref{eq:PhivThetav_ex}, the eigenvalue $\uvec$ of $\uvech$ can take any element of the lattice $\Xi$ 
and thus be expressed as 
\begin{equation}
 \uvec = \sum_{j=1}^4 \sum_{\chlab=1}^2 n_\chlab^{(j)} \avec_\chlab^{(j)}, 
\end{equation}
where $n_\chlab^{(j)}$ is an integer representing the winding number of $\phi_\chlab^{(j)}$ in Fig.~\ref{fig:Z2}(a) in the $x$ direction. 
After imposing the boundary condition $\Gamma_2$, $\uvec$ lives on the reduced lattice $\Xi_{N2}$, 
where as indicated by Eq.~\eqref{eq:tvec}, the winding numbers obey the constraints:
\begin{equation}
\begin{split}
 n_2^{(1)}=n_2^{(2)}, \quad n_1^{(2)}=n_1^{(3)},\\
 n_2^{(3)}=n_2^{(4)}, \quad n_1^{(4)}=n_1^{(1)}.
\end{split}
\end{equation}
These equations simply mean that the winding numbers of $\phi_\chlab^{(j)}$'s on two sheets connected through $\Gamma_2$ in Fig.~\ref{fig:Z2}(a) should take the same integer. 
Similarly, Eq.~\eqref{eq:svec} implies that 
the winding numbers $m_\chlab^{(j)}$ of $\theta_\chlab^{(j)}$'s on two sheets connected through $\Gamma_2$ should take mutually opposite integers.

%----------------------------
%- Calculation of g_Gamma2
Using the primitive vectors $\{\svec_j\}$, the unit cell volume of $\Xit_{D2}$ is calculated as
\begin{equation}
\begin{split}
 v_0(\Xit_{D2})^2 &= 
 \begin{vmatrix}
  \svec_1\cdot\svec_1 & \svec_1\cdot\svec_2 & \svec_1\cdot\svec_3 & \svec_1\cdot\svec_4 \\
  \svec_2\cdot\svec_1 & \svec_2\cdot\svec_2 & \svec_2\cdot\svec_3 & \svec_2\cdot\svec_4 \\
  \svec_3\cdot\svec_1 & \svec_3\cdot\svec_2 & \svec_3\cdot\svec_3 & \svec_3\cdot\svec_4 \\
  \svec_4\cdot\svec_1 & \svec_4\cdot\svec_2 & \svec_4\cdot\svec_3 & \svec_4\cdot\svec_4 
 \end{vmatrix} \\
 &= \det \left( 2\rt^2 K_+^{1/2} K_-^{1/2} M_2 \right),
\end{split}
\end{equation} 
where $M_2$ is the $4\times 4$ matrix defined in Eq.~\eqref{eq:Mn}. 
Since the general case of integer $n\ge 1$ can be handled by simply replacing $M_2$ 
by the $2n\times 2n$ matrix $M_n$ [defined in Eq.~\eqref{eq:Mn}], 
we proceed our discussion in this general case. 
Now the boundary conditions $\Gamma_{1,2}$ are imposed on a $4n$-component boson. 
We obtain
\begin{equation}
 v_0(\Xit_{D2}) = \left( 2\rt^2 K_+^{1/2} K_-^{1/2} \right)^n (\det M_n)^{1/2}. 
\end{equation}
Similarly, we obtain the unit cell volume of $\Xi_{N2}$ as 
\begin{equation}
 v_0(\Xi_{N2}) = \left( 2r^2 K_+^{-1/2} K_-^{-1/2} \right)^n (\det M_n)^{1/2}. 
\end{equation}
Therefore, using Eq.~\eqref{eq:g_Gamma}, the factor $g_{\Gamma_2}$ is calculated as
\begin{equation}
 g_{\Gamma_2} = \pi^n v_0 (\Xit_{D2})^{1/2} v_0(\Xi_{N2})^{1/2} = (\det M_n)^{1/2}. 
\end{equation}
A similar procedure for $\Gamma_1$ yields $g_{\Gamma_1}=1$.

%************************************************
\subsection{Calculation of reduced density matrix moments} \label{sec:ZGG}
%************************************************

%----------------------------
%- Transition amplitudes
We consider the transition amplitudes,  $Z_{\Gamma_1\Gamma_2}$ and $Z_{\Gamma_1\Gamma_1}$. 
The calculation of $Z_{\Gamma_1\Gamma_2}$ for arbitrary $\beta$ is a difficult issue 
because the $\Rcal$ matrices for the two boundary conditions do not commute with each other.
However, as mentioned at the end of Sec.~\ref{sec:boundary_state}, 
one can still derive the asymptotic expressions in the limit $\beta \gg L \gg 1$~(i.e., $q\to 0$). 
The results are
\begin{align}
 &Z_{\Gamma_1\Gamma_2} \approx \bracket{\Gamma_1}{\Psi} q^{-4n/24} \bracket{\Psi}{\Gamma_2} = q^{-4n/24} g_{\Gamma_1} g_{\Gamma_2}, \\
 &Z_{\Gamma_1\Gamma_1} \approx \bracket{\Gamma_1}{\Psi} q^{-4n/24} \bracket{\Psi}{\Gamma_1} = q^{-4n/24} g_{\Gamma_1}^2, 
\end{align}
from which we obtain
\begin{equation}
 \Tr \rho_A^n \approx  \frac{g_{\Gamma_2}}{g_{\Gamma_1}} 
 = (\det M_n)^{1/2}. 
\end{equation}
This constant is exactly the same with that appearing in Eq.~\eqref{eq:rhon_lambda}. 
So far, we have been concerned only with the regulated part of $\Tr~\rho_A^n$ 
and have neglected divergent contributions from the short-range cutoff. 
In general, the logarithm of the partition function, $\log Z_{\Gamma\Gamma'}$, for a cylinder 
contains terms proportional to the area $\frac{\beta}2 L$ and the circumference $L$ (Refs.~\onlinecite{Cardy88,Hsu10}). 
The coefficient of the circumference term depends on the details of the boundary conditions 
while that of the area term depends only on the bulk properties. 
Therefore, in $-\log (\Tr~\rho_A^n)=-\log (Z_{\Gamma_1\Gamma_2}/Z_{\Gamma_1\Gamma_1})$, 
the area terms cancel out while the circumference terms do not, 
leaving a contribution $\alpha L$.  
In this way, the linear contribution in $S_n$ (with integer $n\ge 2$) found in Sec.~\ref{sec:Renyi_ent_expressions}
is also reproduced.

%************************************************
\section{Numerical analysis}  \label{sec:numerics}
%************************************************

%############################
\begin{figure}
\begin{center}
\includegraphics[width=0.48\textwidth]{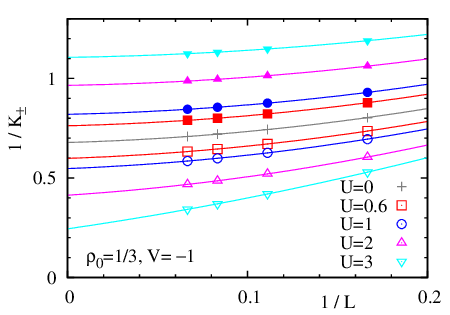}%{KpKm_L_Jlzm0_5.eps}
\end{center}
\caption{(Color online) 
$1/K_\pm(L)$ [Eq.~\eqref{eq:Kpm_L}] versus $1/L$ for $\rho_0=1/3$ and $V=-1$. 
Filled and empty symbols show the data of $1/K_+(L)$ and $1/K_-(L)$ (with $L=6,9,12,15$), respectively. 
Lines show the fitting with the quadratic form 
$1/K_\pm(L)=1/K_\pm + a/L + b/L^2$. 
Our motivation to plot $1/K_\pm(L)$ instead of $K_\pm(L)$ is that 
the formers vary in a smaller range $[0,2]$ in the parameter range of our interest.
}
\label{fig:KpKm_L}
\end{figure}
%############################

%----------------------------
%- Model
In this section, we test the analytical predictions of Secs.~\ref{sec:wavefn} and \ref{sec:BCFT}  
in a numerical diagonalization analysis of a hard-core bosonic model on a ladder. 
The Hamiltonian of the ladder model is given by
\begin{equation}\label{eq:H_boson}
\begin{split}
 H =& \sum_{\chlab =1,2} \sum_{j=1}^L 
 \bigg[ -t\left(b_{j,\chlab}^\dagger b_{j+1,\chlab} + {\rm h.c.}\right) \\
        &+ V \left(n_{j,\chlab}-\frac12\right)\left(n_{j+1,\chlab}-\frac12\right) 
        -\mu \left(n_{j,\chlab}-\frac12\right) \bigg] \\
 +& \sum_{j=1}^L U \left(n_{j,1}-\frac12\right) \left(n_{j,2}-\frac12\right),
\end{split}
\end{equation}
where $b_{j,\chlab}$ is a bosonic annihilation operator at the site $j$ on the $\chlab$-th leg, 
and $n_{j,\chlab}=b_{j,\chlab}^\dagger b_{j,\chlab}$ is the number operator defined from it. 
Here, $t$ and $V$ represent the hopping amplitude and the interaction between nearest-neighbor sites on each leg, 
and $U$ represents the interaction along a rung. 
We impose the hard-core constraint  
$b_{j,\chlab}^2 = (b_{j,\chlab}^\dagger)^2=0$, 
and therefore the bosonic operators are equivalent to spin-$\frac12$ operators as 
$b_{j,\chlab}=S_{j,\chlab}^-,b_{j,\chlab}^\dagger=S_{j,\chlab}^+$. 
%Under this correspondence, the model \eqref{eq:H_boson} is equivalent to 
%two spin-$\frac12$ XXZ chains in a magnetic field, coupled via an Ising interaction. 
We assume the PBC $b_{L+1,\chlab} \equiv b_{1,\chlab}$. 
We define the average particle density as $\rho_0 = (N_1+N_2)/(2L)$, 
where $N_\chlab$ is the particle number on the $\chlab$-th leg. 
We assume $t>0$, $-2< V\le 0$, and $U\ge 0$;  
this case was studied recently in Ref.~\onlinecite{Takayoshi10}. 
We set $t=1$ in the following. 
As explained in Appendix~\ref{app:JW_trans} and in Ref.~\onlinecite{Takayoshi10}, 
this model is equivalent to a fermionic model on a ladder under the Jordan-Wigner transformation. 
In particular, for $V=0$, the model is equivalent to the $SU(2)$-symmetric fermionic Hubbard chain, 
which is solvable by Bethe ansatz. 
In the Hubbard chain, the two legs $\nu=1,2$ are identified with the spin-up/down states, 
and the symmetric/antisymmetric sectors correspond to charge and spin modes, respectively. 

%############################
\begin{figure}
\begin{center}
\includegraphics[width=0.48\textwidth]{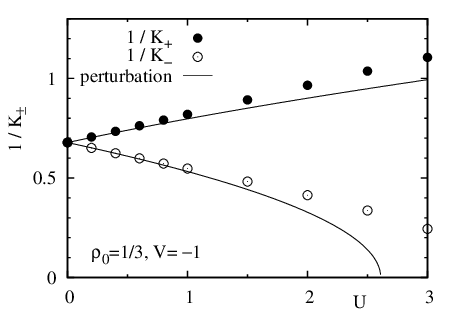}%{KpKm_Jlzm0_5.eps}
\end{center}
\caption{
$1/K_\pm$ for $\rho_0=1/3$ and $V=-1$, 
obtained by extrapolating finite-size data as in Fig.~\ref{fig:KpKm_L}.  
Lines show the perturbative estimates \eqref{eq:vpm_Kpm}.
}
\label{fig:KpKm_U}
\end{figure}
%############################

%----------------------------
%- Result of Takayoshi et al.
We briefly review the recent results of Ref.~\onlinecite{Takayoshi10} on the model \eqref{eq:H_boson}. 
For $U=0$, the model decouples into two independent Bose gases, 
each equivalent to a solvable spin-$\frac12$ XXZ chain in a magnetic field. 
Each chain forms a TLL described by the Hamiltonian \eqref{eq:H_1_2}. 
The velocity $v$ and the TLL parameter $K$ of each XXZ chain can be determined from Bethe ansatz.\cite{Bogoliubov86,Cabra98}
For small $U>0$ and $\rho_0 \ne 1/2$, the inter-chain coupling can be analyzed along the same argument as Sec.~\ref{sec:CTLL}, 
leading to the perturbative estimates \eqref{eq:vpm_Kpm} of the renormalized velocities $v_\pm$ and TLL parameters $K_\pm$ 
(here, the lattice constant is set to unity). 
As seen in this estimate, $K_-$ increases with increasing $U$. 
For $V<0$, it was found that $K_-$ finally diverges as $U$ approaches certain $U_c$, 
where a first-order phase transition to a population-imbalanced state ($N_1\ne N_2$) occurs. 
Here we focus on the uniform phase ($N_1=N_2$) in $0\le U<U_c$ described by the effective Hamiltonian in Eqs.~\eqref{eq:H+H-} and \eqref{eq:H+-}.
In the solvable case $V=0$, the transition is known not to occur, and the uniform phase continues for arbitrary large $U>0$.
In our calculation presented below, we fixed the density at $\rho_0=1/3$, 
and performed calculations for $V=-1,-0.5$, and $0$. 

%############################
\begin{figure}
\begin{center}
\includegraphics[width=0.48\textwidth]{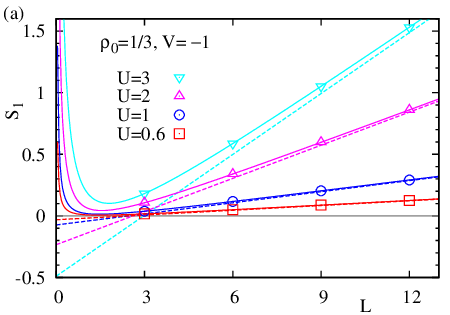}\\
%{ent_L_f2o3Jlzm0_5n1.eps}
\includegraphics[width=0.48\textwidth]{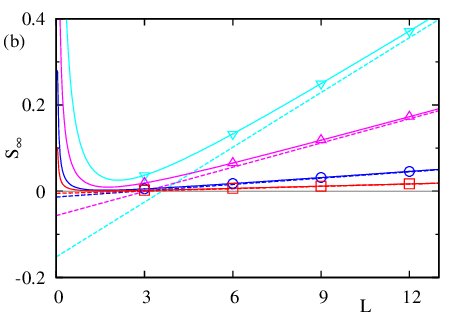}
%{ent_L_f2o3Jlzm0_5ni.eps}
\end{center}
\caption{(Color online)
$S_1$ and $S_\infty$ versus $L$ for $\rho_0=1/3$ and $V=-1$. 
Solid lines show the fits with the scaling form \eqref{eq:Sn_scale} 
and broken lines show the linear part $\alpha_n L+\gamma_n$. 
}
\label{fig:ent_L_Vm1}
\end{figure}
%############################

%----------------------------
%- Estimation of the TLL parameters
Before presenting our results on entanglement, 
let us explain our method for calculating the TLL parameters $K_\pm$. 
In the solvable case $V=0$, $K_\pm$ can be determined accurately by numerically solving the integral 
equations obtained from Bethe ansatz.\cite{Schulz90,Kawakami90,Frahm90} 
For other cases, we determined $K_\pm$ in numerical diagonalization of finite systems (up to $L=15$)  
by using the method of Refs.~\onlinecite{Noack99,Daul98}. 
In this method, we define $\nt_{j,\pm}:=\nt_{j,1}\pm\nt_{2,j}$ with $\nt_{j,\chlab}:=n_{j,\chlab}-\rho_0$, 
and examine their correlation functions 
$C_\pm (r) := \langle \nt_{j,\pm} \nt_{j+r,\pm} \rangle$. 
Using the bosonic representation of operators, 
these correlation functions are shown to have the asymptotic forms
\begin{equation}
 C_\pm (r) = -\frac{K_\pm}{(\pi r)^2} +\frac{A_\pm}{r^{1+K_\pm}} \cos (2k_F r)+\dots,
\end{equation}
where $k_F:=\pi\rho_0$ is the Fermi momentum in the corresponding fermionic model 
and $A_\pm$ are non-universal coefficients. 
In the $SU(2)$-symmetric case, a marginally irrelevant perturbation 
produces multiplicative logarithmic corrections in the second term.\cite{Giamarchi04,Giamarchi89,Schulz90} 
Performing the Fourier transform, 
only the first term contribute for a small wave vector $q$, 
leading to 
\begin{equation}
 N_\pm (q) := \sum_r C_\pm(r) e^{-iqr} \approx \frac{K_\pm}\pi |q| ~~
 (q\approx 0).
\end{equation}
In a periodic finite-size system of length $L$, we evaluate this for $q=2\pi/L$, 
leading to the finite-size estimate of the TLL parameters:
\begin{equation} \label{eq:Kpm_L}
 K_\pm(L) = \frac{L}2 N_\pm \left( \frac{2\pi}{L} \right) .
\end{equation}
The data of $1/K_\pm(L)$ are extrapolated into $L\to\infty$ 
as illustrated in Fig.~\ref{fig:KpKm_L}.
The values of $1/K_\pm$ obtained by the extrapolation are plotted as a function of $U$ in Fig.~\ref{fig:KpKm_U}, 
in reasonable agreement with the perturbative estimates \eqref{eq:vpm_Kpm} for small $U(\lesssim 1)$.

%############################
\begin{figure}
\begin{center}
\includegraphics[width=0.48\textwidth]{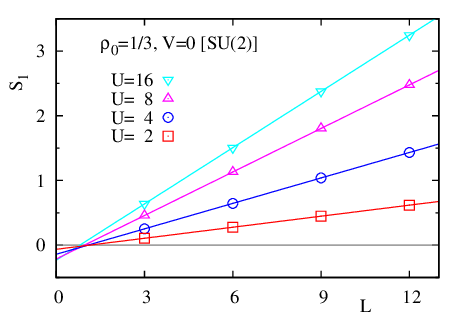}%{ent_L_f2o3Jlz0n1.eps}
\end{center}
\caption{(Color online)
$S_1$ versus $L$ for $\rho_0=1/3$ and $V=0$. 
Solid lines show the fits with the linear form $S_n=\alpha_n L +\gamma_n$. 
}
\label{fig:ent_L_V0}
\end{figure}
%############################

%############################
\begin{figure}
\begin{center}
\includegraphics[width=0.48\textwidth]{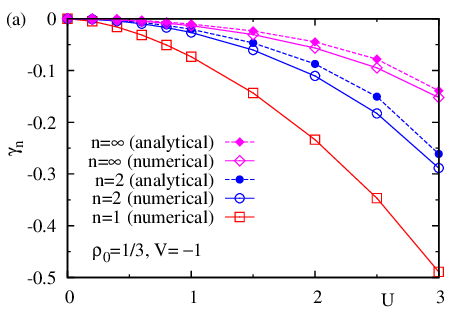}\\
%{entcf2o3Jlzm0_5.eps}
\includegraphics[width=0.48\textwidth]{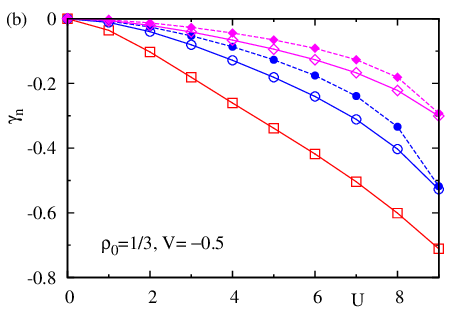}\\
%{entcf2o3Jlzm0_25.eps}
\includegraphics[width=0.48\textwidth]{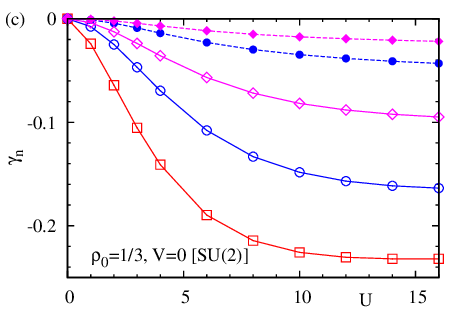}
%{entcf2o3.eps}
\end{center}
\caption{(Color online) 
$\gamma_n$ (with $n=1,2,\infty$) as a function of $U$, 
for $\rho_0=1/3$ and $V=-1,-0.5,0$. 
The analytical formulae of $\gamma_2$ and $\gamma_\infty$ in Eqs.~\eqref{eq:gamma_2} and \eqref{eq:gamma_inf} are also plotted 
using the values of the TLL parameters $K_\pm$ obtained numerically. 
}
\label{fig:entc_U}
\end{figure}
%############################

%----------------------------
%- Entanglement entropy
Let us now present our results on 
the (R\'enyi) entanglement entropies $S_n$ (with $n=1,2,\infty$) between the two legs of the ladder. 
These entropies are calculated 
in the ground states of finite-size systems (up to $L=12$) obtained by Lanczos diagonalization. 
The data of $S_n$ well obey a linear function of $L$. 
For $V=-1$ and $-0.5$, we find that a scaling form\cite{entropy_correction} 
\begin{equation}\label{eq:Sn_scale}
 S_n = \alpha_n L + \gamma_n + \frac{\delta_n}{L} 
\end{equation}
fits the data very well as shown in Fig.~\ref{fig:ent_L_Vm1}. 
%(A similar scaling form was also used in Ref.~\onlinecite{Stephan09}.) 
The linear part $\alpha_n L +\gamma_n$ (broken lines) crosses zero around $L=3$, 
which means that the short-range cutoff $a_0$ discussed in Sec.~\ref{sec:rhon_wavefn_cal} is given by $a_0\approx 3$. 
For $V=0$, in contrast, a simple linear form $S_n=\alpha_n L +\gamma_n$ fits the data better 
as shown in Fig.~\ref{fig:ent_L_V0}. 
The extracted constant $\gamma_n$ is plotted as a function of $U$ in Fig.~\ref{fig:entc_U}. 
Using the values of the TLL parameters $K_\pm$ obtained numerically, 
the formulae of $\gamma_2$ and $\gamma_\infty$ in Eqs.~\eqref{eq:gamma_2} and \eqref{eq:gamma_inf} are also plotted. 
For $V=-1$ and $-0.5$ [Fig.~\ref{fig:entc_U}(a), (b)], we find a broad agreement between the numerical data and the analytical formulae. 
The difference between them are within $\approx 30\%$ of their values. 
We note that our calculations of both $\gamma_n$ and $K_\pm$ are based on finite-size systems with $L\le 15$. 
We expect that calculations in larger systems (by using, e.g., the quantum Monte Carlo method of Ref.~\onlinecite{Hastings10}) 
would demonstrate a more accurate agreement with the analytical predictions. 
For $V=0$ (the Hubbard chain case), on the other hand, we find a significant difference between the numerical and analytical results ---  
the numerical results are roughly four times as large as the analytical results.  
The origin of this significant difference occurring only for $V=0$ will be discussed later in this section.

%############################
\begin{figure}
\begin{center}
\includegraphics[width=0.48\textwidth]{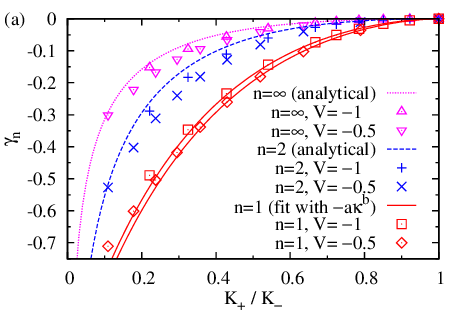}\\
%{entc_KpKm.eps}
\includegraphics[width=0.48\textwidth]{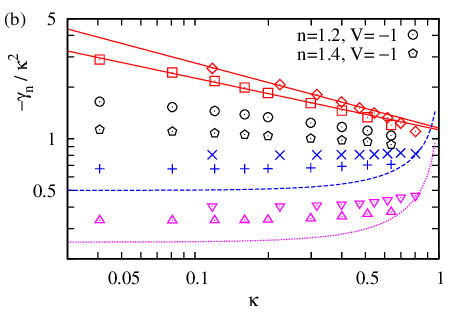}
%{entc_kappa_log.eps}
\end{center}
\caption{(Color online) 
(a) $\gamma_n$ versus $K_+/K_-$. 
(b) $-\gamma_n/\kappa^2$ versus $\kappa:=(K_--K_+)/(K_-+K_+)$ in logarithmic scales. 
The data of $\gamma_1$ are fitted with the form \eqref{eq:gamma1_kappa}. 
The analytical formulae of $\gamma_2$ and $\gamma_\infty$ are from Eqs.~\eqref{eq:gamma_2} and \eqref{eq:gamma_inf}. 
}
\label{fig:entc_KpKm}
\end{figure}
%############################

%----------------------------
%- gamma_n versus K+/K-
In Fig.~\ref{fig:entc_KpKm}(a), we plot the relation of $\gamma_n$ and $K_+/K_-$ using the data for $V=-1$ and $V=-0.5$. 
We can again confirm that for $\gamma_2$ and $\gamma_\infty$, the numerical data and the analytical formulae 
show a broad agreement. 
Furthermore, we observe that the data of $\gamma_1$ for two values of $V$ show a broad agreement, 
which suggests a universal relation between $\gamma_1$ and $K_+/K_-$.  
In Sec.~\ref{sec:Renyi_ent_expressions}, we have expanded $\gamma_n$ (with $n=2,3,\dots$) in terms of $\kappa:=(K_--K_+)/(K_-+K_+)$ 
and found the leading dependence \eqref{eq:gamman_kappa2}. 
Motivated by this observation, we plot $-\gamma_n/\kappa^2$ as a function of $\kappa$ 
in logarithmic scales in Fig.~\ref{fig:entc_KpKm}(b). 
This figure also presents some data for $1<n<2$ obtained in a similar way. 
As expected, the data for $n=2$ and $\infty$ stay around constants as $\kappa$ decreases, 
although these constants are slightly larger than those expected from Eq.~\eqref{eq:gamman_kappa2}. 
The data for $n=1$, however, increase as $\kappa$ decreases, 
and follow straight lines in logarithmic scales in Fig~\ref{fig:entc_KpKm}(b). 
We fit the data with the form $\gamma_1 = -a \kappa^b$ (as mentioned in Eq.~\eqref{eq:gamma1_kappa})
in the range $0<\kappa<0.5$, obtaining $(a,b)\approx(1.13,1.70)$ and $(1.16,1.62)$ for $V=-1$ and $0.5$, respectively. 
This indicates that the leading $\kappa$-dependence of $\gamma_1$ contains a non-trivial exponent $b\approx 1.6$-$1.7$, 
in marked contrast to the quadratic dependence \eqref{eq:gamman_kappa2} of $\gamma_n$ for integer $n\ge 2$. 

In spite of the qualitatively different small-$\kappa$ behaviors for $n=1$ and $n\ge 2$, 
we have found that for fixed $\kappa$, $\gamma_n$ changes rather smoothly when $n$ is changed from $2$ to $1$. 
One can see in Fig.~\ref{fig:entc_KpKm}(b) that the data for $n=1.2$ and $1.4$ indeed intervene between the data of $n=1$ and $2$. 
The issue of how the small-$\kappa$ behavior of $\gamma_n$ changes in the range $1\le n \le 2$ is subtle within the present data. 
We here propose two possible scenarios and leave the issue open for future studies. 
One scenario is that the exponent $b$ decreases smoothly in the range $1\le n \le 2$ 
although it is fixed at $b=2$ for $n\ge 2$. 
Another scenario is that the quadratic behavior of Eq.~\eqref{eq:gamman_kappa2} holds for arbitrary $n>1$, 
but the range of $\kappa$ where the quadratic term dominates 
shrinks gradually as $n$ approaches $1$. 
% Within our current numerical accuracy in estimating $\gamma_n$, 
% it is not easy to discuss which scenario is correct. 

%----------------------------
%- SU(2) case
Finally, let us discuss the origin of the significant difference between numerical and analytical results 
observed for $V=0$ (the Hubbard chain case) in Fig.~\ref{fig:entc_U}(c). 
In $SU(2)$-symmetric systems like the Hubbard chain, 
it is known that a marginally irrelevant perturbation 
produces non-trivial corrections to the predictions of the pure Gaussian model in various physical quantities. 
In particular, its effects are enhanced in the presence of non-trivial boundary conditions, 
as discussed in the spin-$\frac12$ Heisenberg chain\cite{Fujimoto04,Furusaki04,Asakawa96_1} 
and the Hubbard chain\cite{Asakawa96_1,Asakawa96_2,Bortz06} with open ends. 
In the present case, the system has a simple periodic boundary condition in space, 
and non-trivial boundary conditions are imposed in the imaginary time direction 
as presented in Sec.~\ref{sec:BCFT}. 
We expect that a perturbative calculation using boundary states, 
as was done in Ref.~\onlinecite{Fujimoto04}, would clarify non-trivial effects 
of the marginally irrelevant perturbation. 

%************************************************
\section{Summary and discussions}  \label{sec:conclusions}
%************************************************

%----------------------------
%- Summary
We have considered two coupled TLLs on parallel chains 
and calculated the R\'enyi entanglement entropy $S_n$ between the two chains. 
We formulated the problem in the path integral formalism, 
and related $S_n$ with integer $n\ge 2$ to the partition functions on certain non-trivial manifolds. 
These partition functions were calculated using two analytical methods. 
We argued that $S_n$ obeys a linear function of the chain length $L$ 
followed by a universal subleading constant $\gamma_n$. 
The two methods led to the same formulae for $\gamma_n$, 
which are written as functions of the ratio of TLL parameters. 
The obtained formulae were checked numerically 
in a hard-core bosonic model on a ladder. 
When the model is away from the $SU(2)$ case, 
the numerical data of $\gamma_2$ and $\gamma_\infty$ showed a broad agreement with analytical formulae. 
The agreement among two analytical approaches and numerical results  
has offered a convincing evidence of the universality of $\gamma_n$ with integer $n\ge 2$. 
Our numerical results also suggested that the subleading constant $\gamma_1$ in $S_1$ is also universal 
and that its leading dependence on $\kappa:=(K_--K_+)/(K_-+K_+)$ obeys a non-trivial power function, 
in contrast to the quadratic dependence of $\gamma_n$ for integer $n\ge 2$. 
In the $SU(2)$-symmetric case,  
the numerical data of $\gamma_2$ and $\gamma_n$ differ significantly from the analytical formulae, 
which indicates a strong effect of a marginally irrelevant perturbation. 

%----------------------------
%- Comment on particle number fluctuations 
Recently, it has been discussed that 
the particle number fluctuations in a subsystem 
show similar scaling behavior to the entanglement entropy in a number of systems.\cite{Song10} 
This is an interesting proposal relating the entanglement entropy to an experimentally observable quantity. 
In our setting of two coupled TLLs, 
particle number fluctuations in a chain are completely absent since the particle number is separately conserved in each chain. 
On the other hand, finite entanglement entropy does exist between the two chains, 
and obeys a linear scaling with the chain length $L$ as we have discussed. 
Therefore, our study offers a counterexample to the similarity of the two quantities. 
We comment that different behaviors of the two quantities have also been discussed 
in the dynamics of fractional quantum Hall states after a local quantum quench.\cite{Hsu09_qunoise}

%----------------------------
%- Comment on 2D sliding TLL
Our formulations for studying two coupled TLLs can be extended to study the entanglement in multicomponent TLLs. 
An exciting possibility is to study the entanglement entropy 
in a sliding Luttinger liquid,\cite{Schulz83,Emery00,Vishwanath01} 
which appears in a 2D array of coupled TLLs. 
To be specific, we define such a system on a torus of length $L_x$ and $L_y$ in two directions. 
Here, TLLs, described by the bosonic fields $\phi_j(x)$ with $j=1,2,\dots,L_y$, 
are running along the $x$ direction 
and are mutually coupled in the $y$ direction. 
Assuming the translational invariance in the $y$ direction, it is natural to introduce the Fourier transform of the bosonic fields in the $y$ directions:
\begin{equation}
 \phi_q (x) = \frac{1}{\sqrt{L_y}} \sum_j e^{-iqj} \phi_j(x) ,
\end{equation}
with $q=2\pi n_y/L_y$ ($n_y=0,1,\dots,L_y-1$).  
In a sliding Luttinger liquid, the total Hamiltonian decouples into independent TLLs, 
each defined for $\phi_q$ with the renormalized TLL parameter $K_q$ and the velocity $v_q$. 
Now we consider dividing the torus into two cylinders of the same size by cutting it along two lines either in the $x$ or $y$ direction. 
Cutting along $x$ is similar to the problem of this paper; 
it can be treated by generalizing the formulation in Sec.~\ref{sec:BCFT} using more complicated ``mixed'' Dirichlet/Neumann boundary conditions. 
It then leads to the linear scaling of the entanglement entropy with $L_x$,  
followed by a subleading constant determined by $L_y$ TLL parameters. 
The coefficient of the linear term can depend on $L_y$, 
but we expect that it converges to a constant for sufficiently large $L_y$ 
because of the short-range character of the correlations in the $y$ directions. 
When we cut the system along $y$, the original bosonic fields $\phi_j(x)$ are cut at the same positions (say, $x=x_1$ and $x_2$) independent of $j$.  
Then the Fourier components $\phi_q(x)$ are also cut at the same positions for all $q$'s. 
Therefore, the entanglement entropy in this case can be treated in the same way as the single-interval entanglement entropy 
in a 1D gapless system with central charge $c=L_y$. 
Using the finite-system formula in the latter case\cite{Calabrese04,Ryu06} and setting $x_2-x_1=L_x/2$, 
we predict a scaling 
\begin{equation}
\begin{split}
 S &= \frac{L_y}3 \log \left( \frac{L_x}\pi \sin \frac{\pi (x_2-x_1)}{L_x} \right) + {\rm const}. \\
   &= \frac{L_y}3 \log{L_x} + {\rm const.} ~. 
\end{split}
\end{equation}
In these ways, the entanglement entropy shows qualitatively different scaling behaviors 
depending on in which direction one cuts the system. 
Such a highly anisotropic character of entanglement is related to the anisotropic correlations in this system, 
and is in marked contrast to non-interacting fermions\cite{Wolf06,Gioev06,Swingle10} and Fermi liquids.\cite{Swingle10_FL}

%************************************************
\acknowledgements
%************************************************

We are grateful to I. Affleck for many useful comments from the early stage of this work,   
and to T.-P. Choy, A. Furusaki, A. Hamma, and E.S. S{\o}rensen for stimulating discussions. 
We thank M. Oshikawa for showing Ref.~\onlinecite{Oshikawa10} to us prior to publication, 
which motivated us to formulate the boundary conformal field theory approach of Sec.~\ref{sec:BCFT}. 
This work was supported by the NSERC of Canada, 
the Canada Research Chair program, 
and the Canadian Institute for Advanced Research.
We thank the Aspen Center for Physics and the Max Planck Institute for the 
Physics of Complex Systems at Dresden for hospitality, where some parts of this
work were done. 
Numerical diagonalization calculations were performed 
using TITPACK ver.~2 developed by H. Nishimori. 

%%%%%%%%%%%%%%%%%%%%%%%%%%%%%%%%%%%%%%%%%%%%%%%%%
\appendix
%%%%%%%%%%%%%%%%%%%%%%%%%%%%%%%%%%%%%%%%%%%%%%%%%

%************************************************
\section{Ground state wave functional of a TLL}  \label{app:wavefn_TLL}
%************************************************

%----------------------------
%- Short introduction
Here we consider a single-component free boson Hamiltonian 
(defined by the fields $\phi(x)$ and $\theta(x)$) with the TLL parameter $K$, 
and derive the expression of the ground-state wave functional $\bracket{\varphi}{\Psi}$.
Such wave functionals have been derived 
by using the path integral,\cite{Fradkin93,Stone94,Fjaerestad08} 
the Schr\"odinger formalism,\cite{Pham99,Cazalilla04,Fjaerestad08} 
and the Calogero-Sutherland wave function.\cite{Fradkin93,Stephan09} 
This problem is also closely related to 
the effective action for the boundary degrees of freedom 
discussed in the context of dissipation problems\cite{Caldeira83}
and impurity problems.\cite{Furusaki93} 
Here we present a simple derivation in the operator formalism. 
Since the winding numbers (zero modes) of the bosonic fields 
are zero in the ground state, 
we ignore them in the following discussion. 

%----------------------------
%- Mode expansion
The field $\phi$ is expanded as 
\begin{equation}\label{eq:phi_expand}
 \phi (x) = \sum_{m=1}^\infty \sqrt{ \frac{K}{4\pi m} } 
 \big[ (a_m^R+a_m^{L\dagger})e^{ik_m x} + (a_m^L+a_m^{R\dagger})e^{-ik_m x} \big].
 %\left( e^{-ik_mx}a_m^L + a^{ik_mx}a_m^R + {\rm h.c.} \right), 
\end{equation}
with $k_m=2\pi m/L$ and 
$[a_m^L,a_{m'}^{L\dagger}]=[a_m^R,a_{m'}^{R\dagger}]=\delta_{mm'}$. 
This is a one-component version of Eq.~\eqref{eq:PhivThetav_ex}. 
The ground state $\ket{\Psi}$ is defined by $a_m^{L/R}\ket{\Psi}=0~(\forall m\in \mathbb{N})$. 
By analogy with the quantum mechanics of a harmonic oscillator, 
we introduce the ``coordinate''  and ``momentum'' operators, $\Xh_m^{L/R}$ and $\Ph_m^{L/R}$, for each mode via 
\begin{equation}
  a_m^{L/R}          = \frac{\Xh_m^{L/R} + i\Ph_m^{L/R}}{\sqrt{2}}, \quad
 (a_m^{L/R})^\dagger = \frac{\Xh_m^{L/R} - i\Ph_m^{L/R}}{\sqrt{2}}.  
\end{equation} 
The hermittian operators $\Xh_m^{L/R}$ and $\Ph_m^{L/R}$ satisfy the canonical commutation relations 
$[\Xh_m^L,\Ph_{m'}^L]=[\Xh_m^R,\Ph_{m'}^R]=i\delta_{mm'}$. 
%Then, Eq.~\eqref{eq:phi_expand} is rewritten as 
%\begin{equation}\label{eq:phi_expand_XP}
% \phi(x) = \sum_{m\ne 0} \sqrt{ \frac{K}{8\pi |m|} } 
% \left[ (\Xh_m^R+\Xh_m^L) + i (\Ph_m^R-\Ph_m^L) \right] e^{ik_m x}. 
%\end{equation}
We further introduce 
\begin{equation} \label{eq:Xpm_Ppm}
 \Xh_{m,\pm} = \frac{\Xh_m^R \pm \Xh_m^L}{\sqrt{2}} , ~~
 \Ph_{m,\pm} = \frac{\Ph_m^R \pm \Ph_m^L}{\sqrt{2}} , ~~ (m>0)
\end{equation} 
which are related to the ``center of mass'' and ``relative'' 
motions of the left/right-moving modes labeled by $m$. 
Then, Eq.~\eqref{eq:phi_expand} is rewritten as 
\begin{equation} \label{eq:phi_expand_XPpm}
\begin{split}
 \phi(x) = \sum_{m=1}^\infty \sqrt{ \frac{K}{4\pi m} } 
 \bigg[ &(\Xh_{m,+}+i\Ph_{m,-})e^{ik_m x} \\ 
 &+ (\Xh_{m,+}-i\Ph_{m,-})e^{-ik_m x}  \bigg].  
\end{split}
\end{equation}
This expression ``diagonalizes'' $\phi(x)$ 
because all $\Xh_{m,+}$'s and $\Ph_{m,-}$'s commute with each other. 

%----------------------------
%- Wave functional (momentum space)
The state $\ket{\varphi}$ is defined by 
\begin{equation}
 \phi(x) \ket{\varphi} = \varphi(x) \ket{\varphi} \quad (0\le x<L). 
\end{equation}
From Eq.~\eqref{eq:phi_expand_XPpm}, 
one can see that $\ket{\varphi}$ is given by a simultaneous eigenstate of $\{ \Xh_{m,+}; \Ph_{m,-} \}_{m>0}$. 
We expand the field configuration $\varphi(x)$ as 
\begin{equation}\label{eq:vphi_expand}
 \varphi(x) = \frac1{\sqrt{L}} \sum_{m=1}^\infty 
  \left( \varphit_m e^{ik_m x} + \varphit_m^* e^{-ik_m x} \right). 
\end{equation}
Then the coefficient $\varphit_m$ is related to the eigenvalues, $X_{m,+}$ and $P_{m,-}$, of $\Xh_{m,+}$ and $\Ph_{m,-}$ as 
\begin{equation}
 X_{m,+} + iP_{m,-} = \sqrt{\frac{2k_m}{K}} \varphit_m.
\end{equation} 
From the solution of a harmonic oscillator, 
the ground state wave function is written in a Gaussian form in terms of $X_{m,+}$'s and $P_{m,-}$'s as 
\begin{equation}
 \bracket{ \{ X_{m,+}; P_{m,-} \} }{ \Psi } \propto \exp \left[ -\frac12 \sum_{m=1}^\infty (X_{m,+}^2 + P_{m,-}^2) \right].
\end{equation} 
The wave function in terms of $\varphit_m$'s is then given by  
\begin{equation}\label{eq:wavefn_vphi_n}
 \bracket{ \{\varphit_m\} }{\Psi} = \frac{1}{\sqrt{\Ncal}} \exp \left( -\frac1K \sum_{m=1}^\infty k_m |\varphit_m|^2 \right). 
\end{equation}
We normalize the wave function such that 
\begin{equation}
 \int \prod_{m=1}^\infty (d\varphit_m d\varphit_m^*) ~|\bracket{ \{\varphit_m\} }{\Psi}|^2 = 1. 
\end{equation}
Then the normalization factor ${\cal N}$ is calculated as
\begin{equation}\label{eq:Ncal}
 \Ncal = \prod_{m=1}^\infty \int  d\varphit_m d\varphit_m^* \exp \left( -\frac{2k_m}K |\varphit_m|^2 \right)
 = \prod_{m=1}^\infty \frac{\pi K}{k_m}. 
\end{equation}

%----------------------------
%- Wave functional (real space)
It is interesting to transform Eq.~\eqref{eq:wavefn_vphi_n} into the real-space representation:
\begin{equation}\label{eq:wavefn_vphi}
 \bracket{\varphi}{\Psi} = \frac{1}{\sqrt{\Ncal}} e^{-\frac1K \Ecal [\varphi]},  
\end{equation}
with 
\begin{equation}\label{eq:E_vphi}
\begin{split}
 \Ecal [\varphi] 
 = - & \frac1{2\pi} \int_0^L dx_1 \int_0^L dx_2 \\
 &\partial_x\varphi(x_1) \partial_x\varphi(x_2) 
 \log \bigg| e^{i\frac{2\pi}{L}x_1}-e^{i\frac{2\pi}{L}x_2} \bigg| \\ 
\end{split}
\end{equation}
Using the charge density measured from the average, $\delta\rho(x)=-\partial_x \varphi(x)/\sqrt{\pi}$, 
this is rewritten as
\begin{equation}\label{eq:E_Coulomb}
 -\frac12 \int_0^L \delta\rho(x_1) dx_1 \int_0^L \delta\rho(x_2) dx_2 ~
 \log \bigg| e^{i\frac{2\pi}{L}x_1}-e^{i\frac{2\pi}{L}x_2} \bigg|. 
\end{equation}
This can be viewed  as the energy of a classical Coulomb gas placed on a unit circle  
with a logarithmic repulsive potential. 
Such a Coulomb gas structure of the ground state wave function 
is directly seen in the Jastraw-type ground states 
of the Calogero-Sutherland model\cite{Calogero69,Sutherland71} 
and the Haldane-Shastry model\cite{Haldane88,Shastry88}. 
More detailed discussions on these connections 
can be found in Refs.~\onlinecite{Fradkin93,Stephan09}.

%************************************************
\section{Jordan-Wigner transformation for a ladder}  \label{app:JW_trans}
%************************************************

%----------------------------
%- 
Under the Jordan-Wigner transformation, 
the hard-core bosonic model in Eq.~\eqref{eq:H_boson} is equivalent to a spinless fermionic model on a ladder, 
where all the bosonic operators $b_{j,\chlab}$ in Eq.~\eqref{eq:H_boson} are replaced by fermionic ones $f_{j,\chlab}$. 
This transformation is defined as 
\begin{align}
 &f_{j,1} = \exp\left[ i\pi \sum_{l=1}^{j-1} n_{l,1} \right] b_{j,1}, \\
 &f_{j,2} = \exp\left[ i\pi \left( \sum_{l=1}^L n_{l,1} + \sum_{l=1}^{j-1} n_{l,2} \right) \right] b_{j,2}, 
\end{align} 
where the ``string'' part runs first along the first leg and then along the second leg. 
In particular, for $V=0$, the model \eqref{eq:H_boson} is equivalent to the solvable fermionic Hubbard chain, 
where the two legs $\chlab=1,2$ are identified with the spin-up/down states. 
Although the Hamiltonian retains the same form under this transformation, 
the boundary condition is transformed in a non-trivial way. 
For example, the PBC $b_{L+1,\chlab} \equiv b_{1,\chlab}$ on the bosons 
corresponds to the boundary condition $f_{L+1,\chlab} \equiv e^{i\pi N_\chlab} f_{1,\chlab}$ on the fermions, 
where $N_\chlab$ is the number of particles on the $\chlab$-th leg. 
Our motivation to consider the bosonic model \eqref{eq:H_boson} instead of the fermionic one 
is that in the uniform phase which we consider here, the bosonic model \eqref{eq:H_boson} with the PBC has a unique ground state, 
irrespective of the chain length $L$ and the total particle number $N=N_1+N_2$. 
On the other hand, for $U=0$, the fermionic model with the PBC has degenerate ground states for some $L$ and $N$. 
Although this degeneracy is split for $U>0$, some irregular size dependence occurs as a remnant of the degeneracy at $U=0$.

%############################
% \begin{figure}
% \begin{center}
% \includegraphics[width=0.50\textwidth]{.eps}
% \end{center}
% \caption{
% }
% \label{fig:}
% \end{figure}
%############################

%%%%%%%%%%%%%%%%%%%%%%%%%%%%%%%%%%%%%%%%%%%%%%%%%
%References
%%%%%%%%%%%%%%%%%%%%%%%%%%%%%%%%%%%%%%%%%%%%%%%%%

%\end{document}
\newpage

%\appendix
%\begin{document}
%%%%%%%%%%%%%%%%%%%%%%%%%%%%%%%%%%%%%%%%%%%%%%%%%
\section{Note added after publication}
%%%%%%%%%%%%%%%%%%%%%%%%%%%%%%%%%%%%%%%%%%%%%%%%%
\newcommand{\Zbb}{\mathbb{Z}}

\begin{figure*}
\begin{center}
\includegraphics[width=0.48\textwidth]{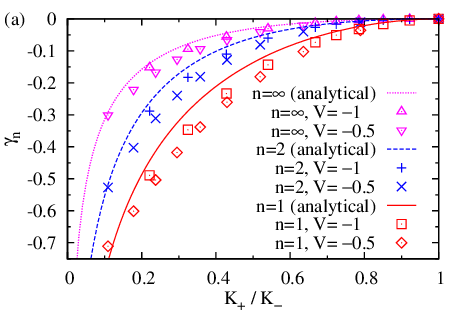}
%{entc_KpKm.eps}
\includegraphics[width=0.48\textwidth]{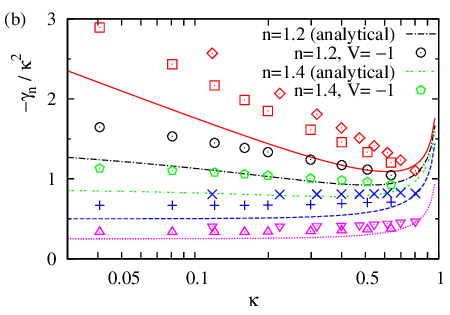}
%{entc_kappa_log.eps}
\end{center}
\caption{(Color online) 
(a) $\gamma_n$ versus $K_+/K_-$. 
(b) $-\gamma_n/\kappa^2$ versus $\kappa:=(K_--K_+)/(K_-+K_+)$ (a logarithmic scale is taken along the horizontal axis). 
These figures replace Fig. 10 of Ref.~\onlinecite{FurukawaKim11}. 
The analytical formulas of $\gamma_n$  are from Eqs.~\eqref{eq:gamma_n_real} and \eqref{eq:gamma_1}. 
}
\label{fig:gamma_n}
\end{figure*}

In our original paper,\cite{FurukawaKim11} 
we derived the formulas of the universal constants $\gamma_n$ for general integer $n\ge 2$, 
and left open the issue of how to analytically continue them to $n\to 1$. 
Here we present a solution to this issue by using a summation trick in Ref.~\onlinecite{Calabrese11}. 
Our solution for general {\it real} $n$ is shown in Eq.~\eqref{eq:gamma_n_real}, 
which is analytically continued to Eq.~\eqref{eq:gamma_1} as $n\to 1$. 
It has turned out that a simple power function [Eq.~(46) of Ref.~\onlinecite{FurukawaKim11}] that we previously assumed for fitting numerical data of $\gamma_1$ was incorrect 
and that the correct small-$\kappa$ behavior of $\gamma_1$ is given by Eq.~\eqref{eq:gamma_1_small_kappa} below. 
We compare the obtained analytic formulas with numerical data in Fig.~\ref{fig:gamma_n}, 
which replaces Fig.~10 of Ref.~\onlinecite{FurukawaKim11}. 

Let us start from Eq.~(39) of Ref.~\onlinecite{FurukawaKim11}, 
which gives $\gammat_n$ for integer $n\ge 1$. 
We aim to extend this to the case of real $n$. 
Differentiating $\gammat_n$ with respect to $\kappa$, we find 
\begin{equation}\label{eq:dgamma_dkappa}
 \frac{d\gammat_n}{d\kappa} = -n \left[ f(0) + f(\pi) \right] + \sum_{l=0}^{2n-1} f\left( \frac{2\pi l}{2n} \right), 
\end{equation}
with 
\begin{equation}
 f(\theta) = \frac{\cos\theta}{1+\kappa\cos\theta}, ~0\le \kappa<1 .
\end{equation}
Since $f(\theta)$ has a period of $2\pi$, we can expand it in a Fourier series: 
\begin{equation}\label{eq:f_Fex}
 f(\theta) = \sum_{k\in \Zbb} f_k e^{ik\theta},~~
 f_k = \int_0^{2\pi} \frac{d\theta}{2\pi} f(\theta) e^{-ik\theta} . 
\end{equation}
The Fourier component $f_k$ can be calculated as follows. 
Introducing $z=e^{i\theta}$, we can rewrite $f_k$ as a contour integral along a unit circle in a complex plane: 
\begin{equation}\label{eq:fk_int}
 f_k = \oint \frac{dz}{2\pi i} \frac{z^{-k-1} (z^2+1)}{\kappa z^2+ 2z + \kappa} .
\end{equation} 
The denominator of the integrand leads to two poles at 
\begin{equation}
 z = z_\pm = \frac{-1 \pm \sqrt{1-\kappa^2}}{\kappa}, 
\end{equation}
which satisfy $z_-<-1<z_+<0$. 
For $k\ge 0$, there is also a pole at $z=0$ coming from the numerator. 
It is sufficient to calculate the integral \eqref{eq:fk_int} for $k\le 0$, 
and then the expression for $k\ge 1$ is obtained by using $f_k=f_{-k}$. 
Consequently, $f_k$ for $k\in\Zbb$ is obtained as
\begin{equation}\label{eq:fk}
 f_k = \frac1\kappa \delta_{k,0} - \frac{ z_+^{|k|} }{ \kappa \sqrt{1-\kappa^2} }. 
\end{equation} 

Using Eqs.~\eqref{eq:f_Fex} and \eqref{eq:fk}, the sum in Eq.~\eqref{eq:dgamma_dkappa} can be calculated as
\begin{equation}
\begin{split}
 &\sum_{l=0}^{2n-1} f\left( \frac{2\pi l}{2n} \right) 
 = \sum_{k\in \Zbb} f_k \sum_{l=0}^{2n-1} e^{ik \frac{2\pi l}{2n} } 
 = 2n \sum_{k\in \Zbb} f_{2nk} \\
 &= 2n f_0 + 4n \sum_{k=1}^\infty f_{2nk}
 = 2n f_0 - \frac{4n}{\kappa\sqrt{1-\kappa^2}} \frac{z_+^{2n}}{1-z_+^{2n}}. 
\end{split}
\end{equation}
It is useful to rewrite the other terms in Eq.~\eqref{eq:dgamma_dkappa} as
\begin{align}
 f(0) &= \sum_{k\in\Zbb} f_k = f_0 + 2\sum_{k=1}^\infty f_k = f_0 - \frac{2}{\kappa\sqrt{1-\kappa^2}} \frac{z_+}{1-z_+}, \\
 f(\pi) &= \sum_{k\in\Zbb} f_k (-1)^k = f_0 + \frac{2}{\kappa\sqrt{1-\kappa^2}} \frac{z_+}{1+z_+}. 
\end{align}
Equation \eqref{eq:dgamma_dkappa} is then calculated as
\begin{equation}
 \frac{d\gammat_n}{d\kappa} = - \frac{4n}{\kappa\sqrt{1-\kappa^2}} 
 \left( \frac{z_+^{2n}}{1-z_+^{2n}} - \frac{z_+^2}{1-z_+^2} \right). 
\end{equation}
Noticing $\frac{dz_+}{d \kappa} = \frac{z_+}{\kappa\sqrt{1-\kappa^2}}$, 
this can be easily integrated, yielding
\begin{equation}
 \gammat_n = 2 \log \left[1-\left(z_+^2\right)^n \right] - 2n \log (1-z_+^2). 
\end{equation}
Although this equation is derived in the case of integer $n\ge 1$, 
it can be directly extended to the case of real $n$. 
We note that $\left( z_+^2 \right)^n$ should not be replaced by $z_+^{2n}$ 
since, given $z_+<0$, the latter does not smoothly depend on $n$. 

%############################
%############################

The universal constants $\gamma_n = -\gammat_n/\left[2(n-1)\right]$ for real $n$ are finally obtained as
\begin{equation} \label{eq:gamma_n_real}
 \gamma_n = -\frac{1}{n-1} \log \left[ 1-\left(z_+^2\right)^n \right] + \frac{n}{n-1} \log (1-z_+^2). 
\end{equation}
The von Neumann limit $n\to 1$ is calculated as
\begin{equation}\label{eq:gamma_1}
\begin{split}
 \gamma_1 &= \log (1-z_+^2) - \lim_{n\to 1} \frac1{n-1} \log \frac{ 1-\left(z_+^2\right)^n}{ 1-z_+^2 } \\
 &= \log (1-z_+^2) + \frac{z_+^2}{1-z_+^2} \log z_+^2
\end{split}
\end{equation}
For $\kappa\ll 1$, we have $z_+\approx -\kappa/2$ and thus
\begin{equation}\label{eq:gamma_1_small_kappa}
 \gamma_1 \approx \frac{\kappa^2}4 \left[ 2 \log \frac{\kappa}{2} -1 \right]. 
\end{equation}
This result indicates that a simple power function [Eq.~(46) in Ref.~\onlinecite{FurukawaKim11}] 
which we previously assumed for fitting numerical data of $\gamma_1$ was incorrect. 

We compare the obtained formulas \eqref{eq:gamma_n_real} and \eqref{eq:gamma_1} with numerical data in Fig.~\ref{fig:gamma_n}. 
The numerical data remain unchanged from Fig.~10 of Ref.~\onlinecite{FurukawaKim11}. 
In Fig.~\ref{fig:gamma_n}(a), we find a broad agreement between numerical and analytical results, 
although the numerical data tend to be slightly smaller than the analytical formulas. 
In Fig.~\ref{fig:gamma_n}(b), the numerical and analytical results show similar qualitative behaviors, 
but we find some appreciable difference particularly for small $\kappa$. 
This indicates a difficulty in correctly obtaining the small-$\kappa$ behavior 
within the system sizes used in our exact diagonalization analysis. 

%############################
% \begin{figure}
% \begin{center}
% \includegraphics[width=0.50\textwidth]{.eps}
% \end{center}
% \caption{
% }
% \label{fig:}
% \end{figure}
%############################

%%%%%%%%%%%%%%%%%%%%%%%%%%%%%%%%%%%%%%%%%%%%%%%%%
%References
%%%%%%%%%%%%%%%%%%%%%%%%%%%%%%%%%%%%%%%%%%%%%%%%%


\begin{thebibliography}{99}
\bibitem{comment_compactification}
The present conditions, in fact, correspond to a bosonic system. 
In a fermionic system, the winding numbers obey the following conditions, 
called the ``twisted structure''\cite{Haldane81_PRL,Wong94,Oshikawa06}:
$n_\chlab \in \mathbb{Z}$, $n_\chlab+2m_\chlab \in 2\mathbb{Z}$. 
Namely, $n_\chlab$ takes an arbitrary integer, 
and $m_\chlab$ take an integer (a half-odd-integer) when $n_\chlab$ is even (odd). 
To simplify the argument, we focus on the bosonic case in the present paper. 
We expect that the bosonic and fermionic cases lead eventually to the same results 
on the ground-state entanglement. 
This is because in the absence of particle tunneling between the chains,
the two cases are related through the Jordan-Wigner transformation 
as discussed in Appendix \ref{app:JW_trans}. 
When $N_1=N_2=$even, a bosonic system with PBC is equivalent to a fermionic system with PBC, 
and thus they show the same ground-state entanglement. 
The differences between the two systems occur in the excitations 
which change $N_1$ and/or $N_2$ to odd.
\bibitem{notation_boson}
Different normalizations are used in the present paper and in Ref.~\onlinecite{Oshikawa10}. 
They can be converted to each other through the correspondences: 
$\Phiv \leftrightarrow \vec{\phi}$, 
$\Thetav \leftrightarrow \vec{\theta}/g$,
$\uvec \leftrightarrow \vec{R}$, 
$\vvec \leftrightarrow \vec{P}=\vec{K}/g$, 
$L \leftrightarrow \beta$, 
$\beta/2 \leftrightarrow l$,  
$q \leftrightarrow \qt$, 
$\qt \leftrightarrow q$, 
where the left and right hand sides correspond to the present paper and Ref.~\onlinecite{Oshikawa10}, respectively. 
Furthermore, $g$ in Ref.~\onlinecite{Oshikawa10} is set to $2\pi$ in the present paper.
\bibitem{entropy_correction}
At present we do not have any derivation of the third term $\delta_n/L$. 
Equation \eqref{eq:Sn_scale} is an empirical form which worked efficiently in Ref.~\onlinecite{Stephan09}. 
In the current problem, we have found that 
the estimations of $\alpha_n$, $\gamma_n$, and $\delta_n$ are rather insensitive 
to which range of $L$ (e.g., $\{3,6,9\}$ and $\{6,9,12\}$) we use for the fitting. 
This lends partial support to this empirical form. 
However, we leave a possibility 
that this simple empirical form does not capture the correct asymptotic behavior 
and leads to a small error in the estimation of $\gamma_n$. 
Further discussion on the correction to the linear form 
is beyond the scope of the current numerical analysis based on small systems.
\end{thebibliography}
\end{document}